\documentclass[12pt]{article}

\usepackage{amsmath}
\usepackage{a4wide}
\usepackage{axodraw}
\usepackage{epsfig}
\usepackage{amssymb}
\usepackage{mathrsfs}

\begin{document}

\begin{titlepage}
  \begin{flushright}
    hep-ph/0406150 \\
    LU TP 04 - 24 \\
    June, 2004
  \end{flushright}
\begin{center}
  {\Huge\bf Saturation In Deep Inelastic Scattering}\\[15mm]
{\large  Master of Science Thesis by Emil Avsar}\\ [2mm]
{\large  Thesis advisor: G\"osta Gustafson}\\ [2mm]
{\it Department of Theoretical Physics,}\\[1mm]
{\it Lund University, Lund, Sweden}
\begin{figure}[b]
\begin{center}
    \includegraphics[angle=0, scale=0.4]{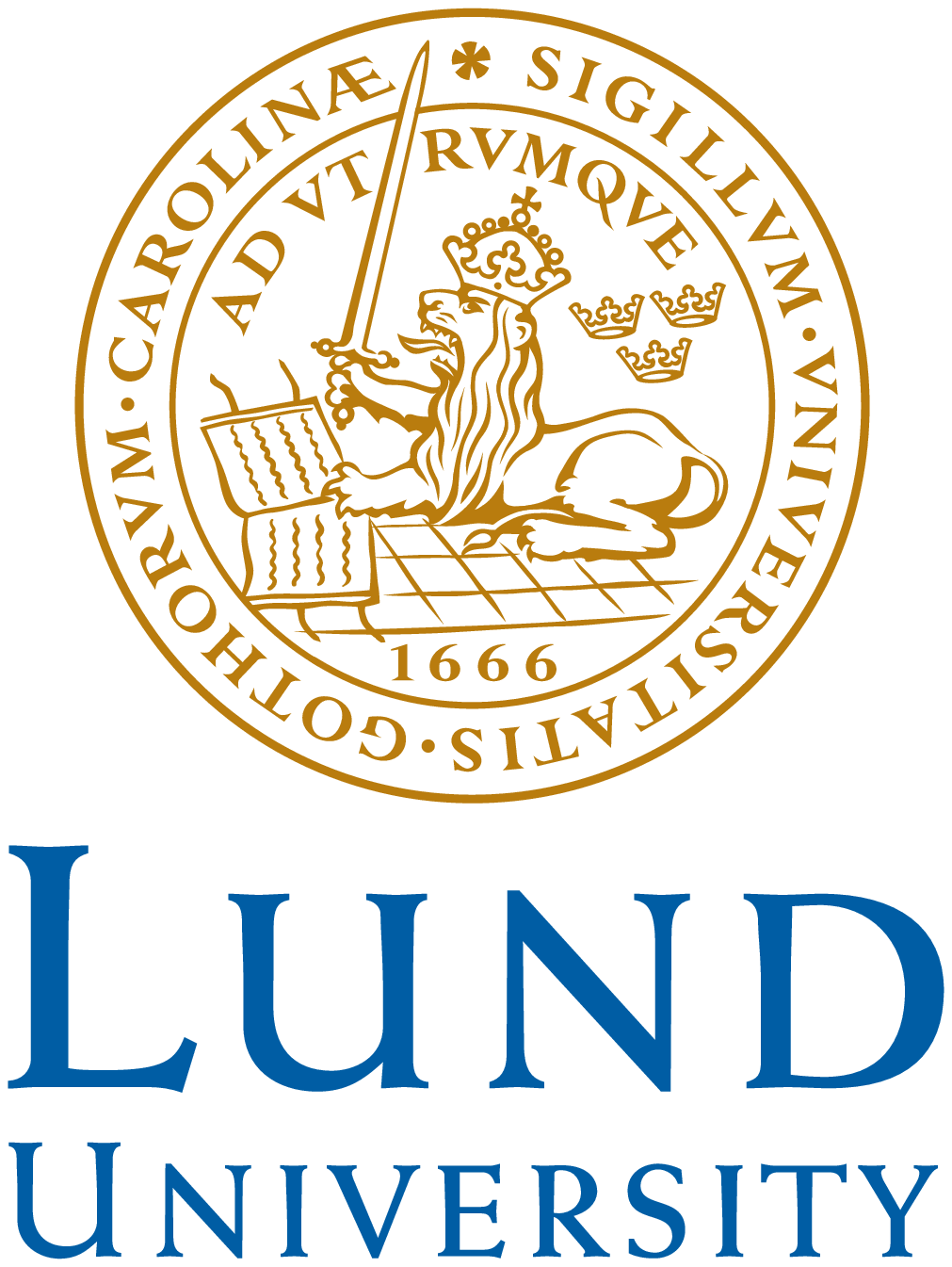}
\end{center}
\end{figure}
\begin{abstract}
The solution to the BFKL equation grows like a power of center of mass energy, $s$, violating unitarity conditions at high energies. The growth of the cross section can be tamed by taking into account multiple pomeron exchanges. This is known as saturation and it is expressed in the Balitsky-Kovchegov equation, \cite{e6}. Conservation of energy should also slow down the growth of the cross section, and our aim in this work is to study the effects of enforcing energy conservation in DIS-events. Using the dipole picture, onium-onium collisions can be viewed, in the large $N_c$ and the leading logarithmic limits, as the scattering of a collection of color dipoles. We construct a Monte Carlo program, based on Mueller's model, \cite{e2, e3, e4, e5}, and using energy conservation, to study onium-onium and onium-nucleus collisions. Dipole fusion processes will be important at high gluon densities and should also slow down the growth of the cross section. We propose an expression for the fusion factor and use it in our Monte Carlo to study its effects.
\end{abstract}
\end{center}
\end{titlepage}

\tableofcontents
\newpage

\section{Introduction}

\begin{quotation}
\emph{It remains that, from the same principles, I now demonstrate the frame of the System of
the World. \\
Isaac Newton \\
Book III, Philosophiae Naturalis Principia Mathematica, 1687}
\end{quotation}

Deep Inelastic Scattering (DIS) is a scattering process where the target particle
disintegrates and what comes out is very different from what came in. If on
the other hand the system after the collision looks exactly like the system
before the collision we call the process elastic. A very common process to
study is the collision between an electron and a proton. This process is usually deeply inelastic at high energies, 
which is due to the fact that the proton is not an elementary
particle but is instead made up of even smaller particles, the quarks. This was
observed in 1968, and it was found that the proton consist of three quarks,
two up-quarks and one down-quark. However, the picture is not as simple as
that. Since the quarks interact via the strong force which is
mediated by the gluons, there are also a lot of gluons in the proton. These gluons can create quark-antiquark pairs or split
into new gluons, and thus the proton structure is much more complicated. It was
indeed observed in these collisions that almost half of the proton's momentum
is carried by particles that don't interact electromagnetically, namely the
gluons. Gluons and quarks are collectively called partons. 

Let us consider a typical electron-proton scattering event, see figure 1. We have an incoming
electron with four-momentum $p$ and an incoming proton with four-momentum
$P$ interacting via the exchange of a virtual photon with virtuality $Q^{2}=-q^{2}$.

\begin{figure}
\begin{center}
\begin{picture}(250,150)(0,0)
\Line(0,0)(200,0)
\Line(0,5)(200,5)
\Line(0,10)(100,10)
\ArrowLine(100,10)(110,50)
\ArrowLine(0,100)(100,100)
\Photon(100,100)(110,50){4}{5}
\ArrowLine(100,100)(200,140)
\ArrowLine(110,50)(200,50)
\Text(50,90)[]{$p$}
\Text(50,20)[]{$P$}
\Text(115,30)[]{$P'$}
\Text(115,80)[]{$q$}
\Text(150,60)[]{$P'+q$}
\end{picture}
\end{center}
\caption 
{\emph{Deep Inelastic scattering between an electron and a proton.}}
\end{figure}
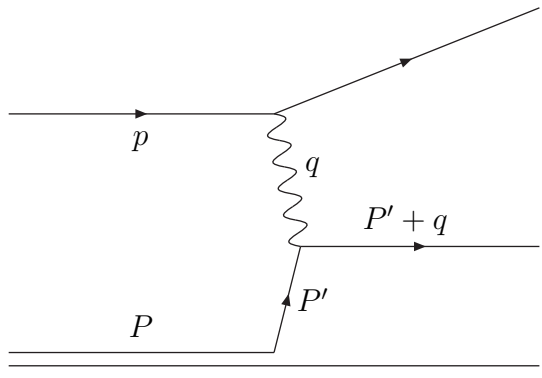

With $P'$ we denote the four-momentum of the parton which
absorbs the virtual photon. The simplest possible process is that it stays on the mass-shell after the absorption,
with a momentum $P'+q$. We define the variable $x$, as the fraction of the proton's momentum
carried by the parton, which is struck by the virtual photon; thus $ x=P'/P$. Since the mass of the proton, and naturally its partons, can be
neglected at the energy scales we are concerned about, we get $ (P'+q)^{2}=0$ and 
$P'^{2}+2P'\cdot q+q^{2}=2xP\cdot q-Q^{2}=0$, and hence $x=\frac{Q^{2}}{2P\cdot
  q}$. This variable $x$ is called the Bjorken scaling variable, usually denoted by
$x_{Bj}$. 

\section{Theory}

We define, the parton distribution function $f(x)$, as the probability for finding a parton carrying a momentum fraction
$x$ of the proton. Usually we use subscripts to denote which parton we are
considering, for example with $f_q(x)$ we denote the probability for finding a
quark with momentum fraction $x$. To specify the flavor of the quark we use
the subscripts u,d etc. The parton model predicts that these distribution functions depend only on the variable $x$. This behavior is known as scaling, which is also why $x$ is called the Bjorken scaling variable. If a parton has a certain $x$ it could either have this momentum fraction
 from the beginning, or it could initially have a larger momentum fraction and reduce it to $x$ through emission of gluons
. The avaliable phase space for radiating gluons is larger when $Q^2$ is large.
Therefore the distribution functions will show a greater probability to have smaller $x$, 
and smaller probability for carrying large $x$ for larger $Q^2$. This implies that the 
distribution functions are not only functions of $x$, but they are functions of
$x$ and $Q^2$, $f(x)=f(x,Q^2)$. This is a central feature of QCD, and it is called scaling violation. 
It is possible to derive evolution equations for the distribution functions in the variable $Q$
 ($\log Q$ is often used instead) if one considers the various processes in the proton, 
such as quarks radiating gluons or gluons splitting into gluons or quark-antiquark pairs. Thus, if one knows the distribution
function for a certain value of $Q$, the value for any other $Q$ can be calculated
using these equations, which in the case of QCD are known as the
\emph{Altarelli-Parisi Equations}. These equations can be found in any book on
Quantum Field Theory.  

The distribution functions are important because the cross sections for different events (both
in QED and QCD) can be expressed in these function. Thus,
for the process we considered above, an electron hitting a proton, the cross
section can be written as 
\begin{equation}
\sigma = \sum_{q,\bar{q}} \int dx[f_q(x,Q^2) \sigma (e^- q \to e^- q) +
f_{\bar{q}}(x,Q^2) \sigma (e^- \bar{q} \to e^- \bar{q})]
\label{eq:e1}
\end{equation}
Here the cross sections in the integrand are the cross sections for the
sub-collisions between the electron and the partons, and the sum runs over the quark and the antiquark flavors.

From now on we will use capital letters to denote the densities in the
variable ln$1/x$, $F(x,Q^2)$ for fermions and $G(x,Q^2)$ for gluons.

\subsection{DGLAP, BFKL and the LDC Models}

\subsubsection{DGLAP evolution}

Consider now the emission of many gluons before the parton is hit by the photon. This is
represented by a fan diagram, see figure 2. 
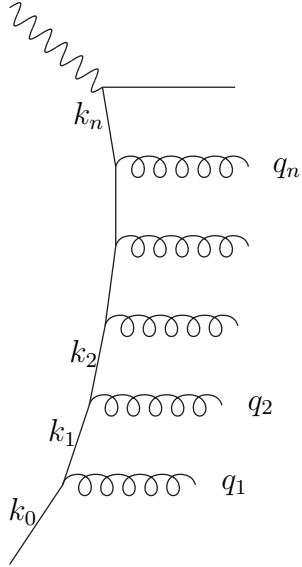
\begin{figure}
\begin{center}
\begin{picture}(200,210)(0,0)
\Line(0,0)(20,30)
\Line(20,30)(30,60)
\Line(30,60)(36,90)
\Line(36,90)(40,120)
\Line(40,120)(40,150)
\Line(40,150)(35,180)
\Photon(0,210)(35,180){4}{6}
\Gluon(20,30)(70,30){4}{5}
\Gluon(30,60)(80,60){4}{5}
\Gluon(36,90)(86,90){4}{5}
\Gluon(40,120)(90,120){4}{5}
\Gluon(40,150)(90,150){4}{5}
\Line(35,180)(85,180)
\Text(5,20)[]{$k_0$}
\Text(20,50)[]{$k_1$}
\Text(28,80)[]{$k_2$}
\Text(30,170)[]{$k_n$}
\Text(85,30)[]{$q_1$}
\Text(95,60)[]{$q_2$}
\Text(105,150)[]{$q_n$}
\end{picture}
\end{center}
\caption
{ \emph{A fan diagram for a DIS event. The virtual propagators
  are denoted by $k_i$ and the quasi-real gluons from the initial-state radiation
  are denoted by $q_i$.}}
\end{figure}
We define $z_i$ as, $z_i=k_{+i}/k_{+i-1}$, hence $1-z_i$ is the fraction
of positive light cone momentum, $p_+=p_0+p_l$, carried away by the quasi-real gluon,
$q_i$. This implies that the Bjorken variable, $x$, is given by $x=\prod_{i=1}^n
z_ix_0$. In the double leading log approximation the gluon ladder is strongly ordered in transverse momentum
\begin{equation}
q_{\perp 1}^{2} < q_{\perp 2}^{2} < \cdots q_{\perp n}^{2} < Q^2
\label{eq:e3}
\end{equation}
and in $x$, $z_i \ll 1$.  
The probability, in the leading log approximation, to emit a gluon is given by 
\begin{equation}
P \sim \frac{4 \alpha _s}{3 \pi} \frac{dz_i}{1-z_i}
\frac{dq_{\perp ,i}^2}{q_{\perp ,i}^2} 
\label{eq:e4}
\end{equation}

This factor must however be modified because the ordering in~\eqref{eq:e3} implies that no gluon is emitted with $q_\perp$ between $q_{\perp,i-1}$ and $q_{\perp,i}$. Thus we must also take into account the probability that no emission has occurred between these two values. This correction factor is called a
Sudakov form factor, and it is, in this case, given by 
\begin{equation}
S(q_{\perp ,i}^2 ,q_{\perp ,i-1}^2 )=exp\biggl[-\int_{q_{\perp ,i-1}^2}^{q_{\perp
    ,i}^2} \frac{4 \alpha _s}{3 \pi} \frac{dq_{\perp ,i}^2}{q_{\perp ,i}^2}
    \int_{0}^{1-\epsilon } \frac{dz_i}{1-z_i} \biggr]
\end{equation}
where the $\epsilon$ is introduced as a cutoff since the integral diverges for
$z=1$. Generally, if we have an ordering in a variable $t$ and the
probability, $P(t)$, for something to occur at a certain $t$ value, the Sudakov form
factor is given by 
\begin{equation}
S(t_1 ,t_2 )=exp\biggl[-\int_{t_1}^{t_2} dtP(t) \biggr]
\end{equation}
Here $(t_1,t_2)$ is the interval where an emission was allowed but did not occur.
Summing over the number, $n$, of links in the chain, the following expression for the
structure function can be derived (for a quark chain)
\begin{equation}
F\sim \sum_{n} \prod_{i=1}^{n} \biggl[ \int \frac{4 \alpha _s}{3
  \pi}  \frac{dq_{\perp ,i}^2}{q_{\perp ,i}^2}  \frac{dz_i}{1-z_i} S(q_{\perp
  ,i}^2 ,q_{\perp ,i-1}^2 ) \theta (q_{\perp ,i} - q_{\perp ,i-1})\biggr] \delta (x -
  \prod_{j}^{n} z_j x_0 ) \theta ( Q^2 - q_{\perp, n}^2) 
\label{eq:dglap}
\end{equation}

Subleading corrections in ln$1/x$ are taken into account by replacing the $1/(1-z_i)$ term by the full
splitting function, $P(z)$. Including spin effects gives
\begin{equation}
\frac{1}{1-z} \longrightarrow \frac{1}{2} \frac{1+z^2}{1-z}
\end{equation}
By taking the derivative of eq.~\eqref{eq:dglap} with respect to
ln$Q^2$ we obtain the DGLAP (Dokshitzer-Gribov-Lipatov-Altarelli-Parisi)
evolution equation 
\begin{equation}
\frac{\partial F(x,Q^2)}{\partial \textrm{ln}Q^2} = \frac{4 \alpha _s}{3 \pi} \int
dzdx'\hat{P}(z)F(x',Q^2)\delta (x-zx')
\end{equation}
Here we have replaced $P$ with $\hat{P}$ (we describe $\hat{P}$ below), and using $\hat{P}$ one can let $\epsilon \rightarrow 0$.  
This can be done since the divergence can be treated by defining a distribution that can be integrated
by subtracting a delta function from the singular term. We define the
distribution, $\frac{1}{(1-z)_+}$, such that for any given smooth function $f(z)$ we have
\begin{equation}
\int_{0}^{1} dz\frac{f(z)}{(1-z)_+} = \int_{0}^{1} dz\frac{f(z) - f(1)}{(1-z)}
\end{equation}
and then we replace the splitting function $P$ with $\hat{P}$ given by
\begin{equation}
\hat{P} = \frac{1}{2} \frac{1+z^2}{(1-z)_+} + \frac{3}{2} \delta (1-z)
\end{equation}

In fact, it is $\hat{P}$ and not $P$ that appears in the AP equations. The
delta function comes from the fact that we have to take into account that a parent
quark splitting into a quark and a gluon
disappears. Thus, we get contributions from the new particles but the one
in the initial state no longer exists, and this must be taken into account. The
coefficient in front of the delta function, $3/2$, is found by the
requirement that the number of quarks is conserved, which means that the
integral of $\hat{P}$ over $z$ must be zero. In the case of gluons (i.e a process where a gluon splits into two gluons) the number of gluons is not conserved but the total energy is, therefore, in this case, the integral of $z\hat{P}$ over $z$ is zero. 

\subsubsection{BFKL region}

Assume that $Q^2$ is only moderately large and that $x$ is kept small. 
The splitting function for the process where a gluon splits into two gluons, $P_{gg}$, contains a 
pole, $1/z$. This implies that for small $x$ and moderately large $Q^2$, this process is very probable 
and will dominate over splitting events involving quarks. Therefore we now consider gluonic chains where a parton emits
a gluon which then develops a chain of gluons by successive splittings into 
new gluons. In this region, non-ordered ladders, though suppressed, will be more important and
they have to be taken into account. This means that the transverse momenta,
$k_\perp$, can now also decrease. Considering this new possibility, Balistky, Fadin, Kuraev and Lipatov derived a more complex
evolution equation, which is known as the BFKL equation. The evolution in $x$
is given by
\begin{equation}
\frac{\partial \mathcal{F}(x,k_{\perp}^2)}{\partial \textrm{ln}(1/x)} = \int dk_{\perp}^{'2}
K(k_{\perp}^2,k_{\perp}^{'2})\mathcal{F}(x,k_\perp^{'2}) 
\label{eq:BFKL}
\end{equation}  
Here we have introduced the unitegrated distribution function, $\mathcal{F}$. There are various definitions of $\mathcal{F}$, and in the LDC model (which is discussed in the next section) it satisfies the following relation 
\begin{equation}
F(x,Q^2)=\int^{Q^2} \frac{dk_{\perp}^2}{k_{\perp}^2} \mathcal{F}(x,k_{\perp}^2) + 
\int_{Q^2} \frac{dk_{\perp}^2}{k_{\perp}^2} \mathcal{F}(x\frac{k_\perp^2}{Q^2},k_{\perp}^2)\frac{Q^2}{k_\perp^2}
\end{equation} 

The kernel of the integral, $K$, in equation~\eqref{eq:BFKL}, is known as the BFKL kernel. The
important thing here is to find the eigenvalues of this kernel. To leading order 
in $\textrm{ln}(1/x)$, and for a fixed coupling
constant $\alpha _S$, the largest eigenvalue is given by $\lambda = 12\alpha _S \log (2)/\pi$. It can then be shown that the gluon distribution function
is given by 
\begin{equation}
G \sim \sum _{n} \frac{(\lambda \log (1/x))^n}{n!} = x^{- \lambda}
\end{equation}

We will come back to the BFKL equation, and the so called BFKL pomeron, later
when we discuss saturation. For now, one can note that BFKL and DGLAP are applicable in
different regions. In the next section we will describe two models which interpolate between these two
different regions, thus containing both the DGLAP and the BFKL physics. 

\subsubsection{CCFM and the LDC Models}

The first such model was proposed by Catani, Ciafaloni, Fiorani and Marchesini
and is thus called the CCFM model. We will discuss purely gluonic chains and
the last gluon in the chain is assumed to interact with a color neutral
probe. One can usually order emissions in different variables, such as color, rapidity, transverse momentum 
and so on. Due to soft color coherence \cite{15} the ordering in rapidity and in color are equivalent however. This is used in the CCFM model where 
the initial state radiation is made up of gluons which are not followed, in rapidity, by a gluon with
more energy, or equivalently positive light-cone momentum, $q_+ =q_0 + q_l$.

With the selection of the initial state radiation described above, it was shown
by the authors that the unintegrated distribution function is given by
\begin{eqnarray}
\mathcal{G}(x,k_{\perp}^2 , \bar{q} ) & \sim & \sum_{n} \prod_{i=1}^{n}
  \biggl[ \int \bar{\alpha }  \frac{dq_{\perp ,i}^2}{q_{\perp ,i}^2} \biggl(
  \frac{1}{z_i} \Delta _{ne} (z_i ,k_{\perp ,i}^2 ,\bar{q_i})+\frac{1}{1-z_i}
  \biggr) \Delta _S \times  \\
  & & \theta (\bar{q_i}-\bar{q}_{i-1}z_{i-1}) \biggr] \delta (x -
  \prod_{j}^{n} z_j ) \theta ( \bar{q}-\bar{q}_n z_n) \delta(k_{\perp}^2
  -k_{\perp n}^2)
\nonumber
\end{eqnarray}
where $\bar{q}_i$ is
defined by $\bar{q}_i \equiv q_{\perp ,i}/(1-z_i)$. $\Delta_{ne}$ stands for a non-eikonal form factor and $\Delta _S$ is
the Sudakov form factor. They are given by
\begin{equation}
\Delta _{ne}=\textrm{exp}
\biggl(-\bar{\alpha}\textrm{ln}\frac{1}{z}\textrm{ln}\frac{k_{\perp}}{z\bar{q}^2}\biggr);
\;\; \Delta _S =\textrm{exp} \biggl(-\bar{\alpha}\int \frac{dq_{\perp ,i}^2}{q_{\perp
    ,i}^2} \frac{dz}{1-z} \Theta _{order} \biggr)
\end{equation}
Here $\Theta _{order}$ specifies the kinematic region allowed by the ordering
constraints, and $\bar{\alpha}\equiv 3\alpha _s/\pi$. In the CCFM model, only the singular
terms, $1/z$ and $1/(1-z)$, in the splitting functions are included. The $1/z$
pole is important in the BFKL region and the $1/(1-z)$ pole and the Sudakov
factor are more important in the DGLAP region. Since the non-singular terms in
the splitting function are not included, DGLAP is not fully reproduced in the
large $Q^2$ and the large $x$ limits. 

The LDC (Linked Dipole Chain) model, \cite{e1},  
 is a reformulation and a generalization of
the CCFM model. The main idea in this model is that certain gluons in the
initial state radiation of the CCFM model, namely those gluons which satisfy
the constraint $q_{\perp i}< \textrm{min}(q_{\perp (i-1)},q_{\perp (i+1)} )$, can be
treated as final state emissions from dipoles, which are formed by the
remaining initial state gluons. These remaining gluons, the initial state
gluons in the LDC model, will then satisfy $q_{\perp i}\approx \textrm{max}(k_{\perp
  i},k_{\perp (i-1)})$. They will be ordered in both positive and negative
light-cone momenta, $q_{-}=q_0-q_l$. A single chain in the LDC model corresponds to several
chains in CCFM, and it was shown in \cite{e1} that the non-eikonal
factors in CCFM add up to one, when one considers all contributions from these
chains. The ordering, in both $p_+$ and $p_-$, and the absence of the non-eikonal form factors,
implies that LDC is forward-backward symmetric. If one starts from the probe
end, which has large $q_{-}$, and follows the chain towards the proton end,
which has large $P_+$, the situation will look the same, the minus and plus
components will just change roles and we will have exactly the same
ordering and the same distribution function. In the LDC model the distribution function satisfies 
the following evolution equation (see eq.43 in \cite{e1}) 
\begin{equation}
\frac{\partial \mathcal{F}(x,k_{\perp}^2)}{\partial \log (1/x)} \approx \bar{\alpha}
\biggl[\int^{k_{\perp}^2}
  \frac{dk_{\perp}^{'2}}{k_{\perp}^{'2}}\mathcal{F}(x,k_{\perp}^{'2}) + \int_{k_{\perp}^2}
  \frac{dk_{\perp}^{'2}}{k_{\perp}^{'2}} \frac{k_{\perp}^{2}}{k_{\perp}^{'2}}
  \mathcal{F}(x\frac{k_{\perp}^{'2}}{k_{\perp}^{2}}, k_{\perp}^{'2}) \biggr]  
\end{equation}
Here $k_{\perp}^{2}$ is the transverse momentum of the last gluon in the chain
while $k_{\perp}^{'2}$ is the momentum in the step before. The first term here is the DGLAP contribution where the transverse momentum
increases in the last step, while the second term contains the probability that
the transverse momentum decreases in the last step. For a more detailed discussion on these models, see \cite{e11}.  

\subsection{Mueller's Dipole Formulation}

The idea of
dipoles as emitters of radiation is quite natural in QCD. Consider the two
diagrams in figure 3. 
\begin{figure}
\begin{center}
\begin{picture}(400,80)(0,0)
\Line(0,80)(30,50)
\Line(0,20)(30,50)
\Photon(30,50)(60,50){3}{4}
\Line(60,50)(90,80)
\Line(60,50)(90,20)
\Gluon(70,60)(110,60){3}{3}
\Line(200,80)(230,50)
\Line(200,20)(230,50)
\Photon(230,50)(260,50){3}{4}
\Line(260,50)(290,80)
\Line(260,50)(290,20)
\Gluon(270,40)(305,40){3}{3}
\Text(65,65)[]{$r$}
\Text(65,35)[]{$\bar{r}$}
\Text(265,65)[]{$b$}
\Text(265,35)[]{$\bar{b}$}
\Text(115,65)[]{$\bar{b}$}
\Text(115,55)[]{$r$}
\Text(85,30)[]{$\bar{r}$}
\Text(90,70)[]{$b$}
\Text(315,45)[]{$\bar{b}$}
\Text(315,35)[]{$r$}
\Text(285,30)[]{$\bar{r}$}
\Text(290,70)[]{$b$}
\end{picture}
\end{center}
\caption
{ \emph{A quark and an antiquark emitting a gluon. In
the first picture the red quark radiates a red-antiblue gluon and changes its
color to blue and in the second diagram the antiblue antiquark emits a
antiblue-red gluon and changes its anticolor to antired.} }
\end{figure}
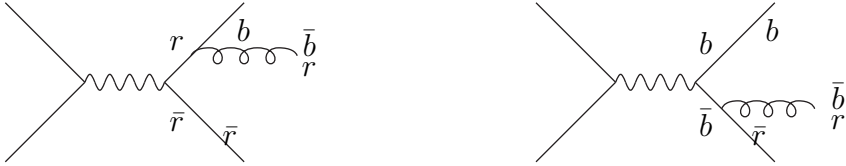
The two diagrams will interfere and the two final states are identical. Thus, it is not possible to tell
whether the quark or the antiquark emitted the gluon. Therefore it
makes sense to regard the quark-antiquark pair as a color dipole, radiating coherently. In the region between the quark and the gluon, the blue and antiblue color separation will give rise to the emission of softer gluons. But in directions further away the emissions from blue and antiblue will interfere destructively. Therefore, in these directions, the emission of softer gluons will correspond to a red antired color dipole. In the restframe of the quark and the gluon however, emission of softer gluons corresponds to a blue antiblue color dipole. Therefore the emission of softer gluons corresponds to two independent color dipoles. This generalizes
and a system of a quark, an antiquark and $n-1$ gluons will correspond, in the
large $N_c$ limit, to a system of $n$ independent color dipoles. Each time a gluon is radiated a dipole 
splits into two new dipoles and the process continues. 
This is used in the Dipole Cascade Model, \cite{13, 14}, which is formulated in momentum space. 

A large part of this thesis is based on Mueller's Dipole model
\cite{e2,e3,e4,e5} and we will be using it in our MC program. In Mueller's model we start with a heavy quark-antiquark pair, an onium
state, and consider successive emissions of softer and softer gluons. The main
idea here is also based on the observation that, in the large $N_c$ limit and
in the leading logarithmic approximation the complete, multi-gluon,
wavefunction of an onium state can be regarded as a collection of dipoles. This is so since, 
as we saw above, each gluon will act like a quark-antiquark
pair. Mueller's approach is formulated in transverse position,
$\bold{x}$, instead of transverse momentum, $\bold{k}$. These two variables are naturally connected by a Fourier transform
\begin{equation}
\psi_{\alpha \beta}^{(0)}(\bold{x}_1,Y)=\int
\frac{d^2\bold{k}_1}{(2\pi)^2}e^{i\bold{k}_1\cdot\bold{x}_1}\psi_{\alpha
  \beta}^{(0)}(\bold{k}_1,Y)
\end{equation}
Here $\psi$ is the light-cone wavefunction (the zero on $\psi$ indicating
that there are 0 gluons present) and $\alpha$ and $\beta$ are heavy
quark and antiquark spinor indices, $\bold{x_0}$ and $\bold{x_1}$ are the
transverse coordinates of the quark and the antiquark respectively. Finally,
$Y$ is the total rapidity range determined by the center of mass energy, $Y\sim$ln$s/M^2$, $M$ being the mass of the onium. We also define
$\bold{x}_{10}=\bold{x}_1-\bold{x}_0$, which means that the transverse size of the
initial dipole is, $x_{10} \equiv \vert \bold{x}_{10}\vert$. The reason we choose to work with transverse coordinates 
instead of transverse momenta is that one can assume the partons to be frozen during emissions and the evolution
 process, which makes the analysis much easier. 
This is justified since the time scales for the processes
that we consider are very short. Denoting the
square of the wave function with $\phi$, Mueller showed that 
\begin{equation}
\phi^{(1)}(\bold{x}_1,Y)=\frac{\bar{\alpha}}{2\pi}\int d^2\bold{x}_2
\int_{0}^{Y}
dy\frac{x_{10}^2}{x_{20}^2x_{12}^2}\phi^{(0)}(\bold{x}_1,Y)
\label{eq:mueller2}
\end{equation}     
where $\bold{x_2}$ is the transverse position of the emitted gluon and $\phi^{(0)}$
and $\phi^{(1)}$ are the squared wave functions for the onia state without and
with a gluon respectively. The factor
$d^2\bold{x}_2\frac{x_{10}^2}{x_{20}^2x_{12}^2}$ plays the same role as the
splitting functions we hade earlier. Observe that the $\bold{x}_2$ integral diverges for small $x_{20}$ or $x_{12}$. This means that the
probability to emit a dipole goes to infinity as the dipole size approaches
zero. To deal with this a cutoff, $\rho$, is introduced, which later can 
be sent to zero and will not appear in any physical quantities. Generally, if we have $n$ dipoles and we consider the emission
of the $n$th gluon, we will get a contribution
\begin{equation}
\sum_{ij}\int d^2\bold{x}_n\frac{x_{ij}^2}{x_{in}^2x_{nj}^2}
\end{equation} 
where $ij$ runs over all the indices for which there exist a dipole. 

\begin{figure}
\begin{center}
\begin{picture}(400,200)(0,0)
\Line(100,60)(100,140)
\Line(100,140)(140,160)
\Line(100,60)(140,160)
\Line(100,60)(160,40)
\Line(140,160)(160,40)
\Vertex(100,140){2}
\Vertex(100,60){2}
\Vertex(140,160){2}
\Vertex(160,40){2}
\Line(140,160)(180,175)
\Line(180,175)(160,40)
\Vertex(180,175){2}
\Text(100,150)[]{$0$}
\Text(100,50)[]{$1$}
\Text(140,170)[]{$2$}
\Text(165,40)[]{$3$}
\Text(190,175)[]{$4$}
\Text(185,116)[]{$\bold{\cdots}$}
\end{picture}
\end{center}
\caption
 { \emph{Creation of new dipoles by means of gluon
  emission. The original quark is denoted by 0 while the antiquark is denoted
  by 1. The original dipole "decays" into two new dipoles when the gluon labeled by 2 is emitted. 
The process continues with the emissions of gluons 3 and 4 etc.} } 
\end{figure}
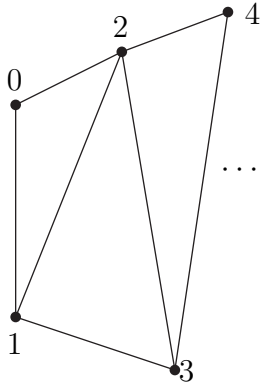

The complete formula for the wave function with arbitrarily many gluons is
derived from a generating functional. Mueller derived a non-linear integral
equation for this generating functional. Taking into account not only real
gluon emissions, but also the leading logarithmic virtual corrections, the
formula reads
\begin{eqnarray}
Z(\bold{b}_0,\bold{x}_{10},Y,u)&=&u(\bold{b}_0,\bold{x}_{10})exp\biggl[-2\bar{\alpha}\textrm{ln}\biggl(\frac{x_{10}}{\rho}\biggr)Y\biggr]+
\nonumber \\
& & +\frac{\bar{\alpha}}{2\pi}\int_0^Y dyexp\biggl[-2\bar{\alpha}\textrm{ln}\biggl(\frac{x_{10}}{\rho}\biggr)(Y-y)\biggr] \times \label{eq:mueller} \\
& & \times \int_\rho d^2\bold{x}_2\frac{x_{10}^2}{x_{20}^2x_{12}^2}Z(\bold{b}_0
  +\frac{1}{2}\bold{x}_{12},\bold{x}_{20},y,u)Z(\bold{b}_0
  -\frac{1}{2}\bold{x}_{20},\bold{x}_{12},y,u)
\nonumber 
\end{eqnarray}
Here the $\rho$ under the second integral indicates that we are integrating
over a region satisfying, $x_{20}\geqslant \rho$ and $x_{21}\geqslant
\rho$. The vector $\bold{b}_0$ is the position of the center of the initial dipole
in transverse coordinates. The two exponentials have the meaning of Sudakov
factors, where in the first one we consider the probability that nothing
happens within the allowed rapidity interval, which means that there are no gluons present. The second exponential contains
the probability that nothing happens after the last gluon is emitted at 
rapidity $y$.

\subsection{Saturation and The Balitsky-Kovchegov Equation } 

Consider the scattering of two onium states against each other. Using the dipole
formulation described above, this process can be viewed as the scattering of
a collection of color dipoles against each other. The single pomeron exchange then corresponds
to the approximation in which two dipoles (from the two different states) interact via the
exchange of a color neutral gluon pair, independently of the rest of
the dipoles. This pomeron is known as the BFKL pomeron. The total cross
section would then be given by the number of dipoles in one onium state
times the number of dipoles in the other onium state times the cross
section for dipole-dipole scattering due to the two gluon exchange. However,
as the center of mass energy increases the number of dipoles will increase,
and when the number becomes sufficiently large the total cross section will violate
the unitarity bounds. The reason is that at higher energies, second order processes
must be taken into account, and these will slow down the growth of the total
cross section. The higher order scatterings are called multiple pomeron
exchanges.  

Consider the scattering of a photon off a nucleus. We assume that the photon
splits into a quark-antiquark pair, on which we apply the dipole formulation, that is we work in the large $N_c$ and the leading logarithmic limits. This means that the photon will develop a cascade of color dipoles, which
interact with the nucleus. The nucleus is assumed to be at rest. We assume that each dipole interacts with the
nucleus by a two gluon exchange, independent from the rest of the dipoles. The nucleus is approximated
by a cylinder which makes the calculations easier. The multiple pomeron
exchanges will correspond to the simultaneous interactions of multiple dipoles
with the nucleus, which means that the $k$:th pomeron exchange corresponds to $k$ dipoles
simultaneously interacting with the nucleus by exchanging $k$ color
neutral gluon pairs. With these considerations
and summing pomeron exchanges up to all orders, Kovchegov, in \cite{e6}, derived
the following non-linear differential-integral equation for the forward
scattering amplitude off the nucleus:
\begin{eqnarray}
\frac{\partial N(\bold{x}_{10},\bold{b}_0,Y)}{\partial
  Y}&=&\frac{\bar{\alpha}}{2\pi}\int_\rho
  d^2\bold{x}_2\frac{x_{10}^2}{x_{20}^2x_{12}^2}\biggl(
  N(\bold{x}_{20},\bold{b}_0 +
  \frac{1}{2}\bold{x}_{12},Y) + \label{eq:BK}\nonumber \\ 
& & +N(\bold{x}_{12},\bold{b}_0 -
  \frac{1}{2}\bold{x}_{20},Y)-N(\bold{x}_{10},\bold{b}_0,Y)- \\
& & -N(\bold{x}_{20},\bold{b}_0 +
  \frac{1}{2}\bold{x}_{12},Y)N(\bold{x}_{12},\bold{b}_0 -
  \frac{1}{2}\bold{x}_{20},Y)\biggr)\nonumber 
\end{eqnarray}

Here the imaginary part of $N$ gives the total cross section. This equation is known as the Balistky-Kovchegov (BK) equation (a very similar 
equation was proposed by Balitsky in \cite{e12}), and as we mentioned above, it
contains all the multiple pomeron exchanges. 
The linear part of the equation is the BFKL part which dominates for small
values of $N$ and which causes the cross section to rise exponentially. The non-linear part
comes from the multiple pomeron exchanges and causes the amplitude to saturate
for values near 1. The equation has a nice physical interpretation, the first two terms on the right hand side are the
contributions to the scattering amplitude from the creation of the new
dipoles, 12 and 20, while the third term comes from the fact that the original
dipole, 10, no longer exists. The reader might guess that the
non-linear term is the contribution from the recombination of the two
dipoles, 12 and 20, into the original dipole 10, but as we said above the
non-linear term comes from the multiple pomeron exchanges. The
Balitsky-Kovchegov equation does not contain the process of two dipoles
recombining into one. This fusion process should also contribute to saturation. 
For large gluon densities it should be non-negligible, but it is
not an easy task to estimate this effect. In fact, in the
derivation of ~\eqref{eq:BK}, Kovchegov used ~\eqref{eq:mueller}, which does not
contain the fusion process of dipoles. We will come back to dipole fusion processes on section 5.

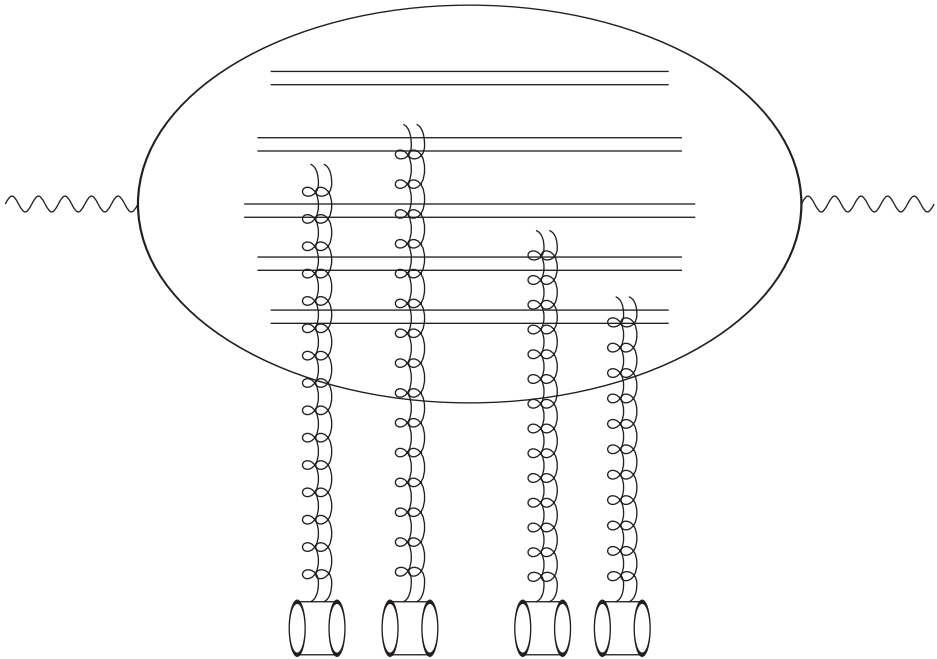
\begin{figure}
\begin{center}
\begin{picture}(400,300)(0,0)
\Photon(0,200)(50,200){3}{5}
\Oval(175,200)(75,125)(360)
\Photon(300,200)(350,200){3}{5}
\Line(100,250)(250,250)
\Line(100,245)(250,245)
\Line(95,225)(255,225)
\Line(95,220)(255,220)
\Line(90,200)(260,200)
\Line(90,195)(260,195)
\Line(95,180)(255,180)
\Line(95,175)(255,175)
\Line(100,160)(250,160)
\Line(100,155)(250,155)
\Gluon(150,230)(150,50){3}{15}
\Gluon(155,230)(155,50){3}{15}
\Gluon(115,215)(115,50){3}{15}
\Gluon(120,215)(120,50){3}{15}
\Gluon(200,190)(200,50){3}{14}
\Gluon(205,190)(205,50){3}{14}
\Gluon(230,165)(230,50){3}{11}
\Gluon(235,165)(235,50){3}{11}
\Oval(145,40)(3,10)(90)
\Oval(160,40)(3,10)(90)
\Oval(110,40)(3,10)(90)
\Oval(125,40)(3,10)(90)
\Oval(195,40)(3,10)(90)
\Oval(210,40)(3,10)(90)
\Oval(225,40)(3,10)(90)
\Oval(240,40)(3,10)(90)
\Line(145,50)(160,50)
\Line(145,30)(160,30)
\Line(110,50)(125,50)
\Line(110,30)(125,30)
\Line(195,50)(210,50)
\Line(195,30)(210,30)
\Line(225,50)(240,50)
\Line(225,30)(240,30)
\end{picture}
\end{center}
\caption
{ \emph{The photon develops a cascade of dipoles which
interact with nucleus by the exchange of color neutral gluon pairs. Each
double line represents a gluon which behaves like a quark-antiquark pair in
the large $N_c$ limit.}} 
\end{figure}

\section{MC Simulation of Onium Evolution With Energy Conservation}

\subsection{The Main Idea}

As we mentioned earlier, the probability to emit small dipoles becomes
infinitely large as the dipole size approaches zero. To deal with this we
introduced a cutoff, $\rho$. This cutoff is an ultraviolet cutoff since a
small dipole corresponds to two hard gluons. This follows from the relation
$\frac{1}{r}\sim p_\perp$, where $r$ is the transverse size of the dipole and $p_\perp$ is
the transerverse momentum. We will now develop an onium state using an
ordering in rapidity, $y$, such that the rapidity increases for each
emission. We find it convenient to define $y=$ln$\frac{p_\perp}{p_+}$, hence $p_+=p_\perp e^{-y}$. We see that, for fixed $y$, 
$p_+\rightarrow \infty$ as $r\rightarrow 0$. This means that, if we allow very
small $r$ we will violate energy conservation since the energy of the gluon
will become larger than the energy of the mother dipole. Hence, we cannot
allow too small dipole sizes. The positive 
light-cone momentum is essentially equal to the energy,
and from now on we will refer to it as energy. Let $p_+$ be the energy of the parent dipole
and let the gluon be emitted at a rapidity
$y$. Denote by $p_{+g}$ the energy of the gluon. We then have
\begin{equation}
p_{+g}\leqslant p_+\Rightarrow p_{\perp g}e^{-y}=\frac{1}{r}e^{-y}\leqslant
p_+, \; \; \textrm{hence}\; r\geqslant \frac{e^{-y}}{p_+}
\end{equation} 

This means that we get a very natural cutoff for small dipoles, given by,
$\rho (y)=e^{-y}/p_+$. To make the program right-left symmetric, which means that the theory looks the same independent of which way we evolve, we must also demand $p_-$ conservation, which is described later. These observations are the main motivations for this thesis. 
We will now describe the Monte Carlo program, including energy
conservation, developed to study the deeply inelastic scattering between a photon and a nucleus, and between two onium states.   

\subsection{Recoil Formulas}

Since we demand energy to be conserved, we must also take into account the
full recoil effects. A single gluon will, in the dipole model, be a part of two independent dipoles
and the question is how  the two
dipoles share the recoil from the emission. We solve this by assuming that when a dipole emits a gluon, the
initial gluons, which make up the parent dipole, may contribute with all their energy. This means
that the next time a neighboring dipole emits a gluon, the avaliable energy is reduced 
because one of its gluons has lost energy from an earlier emission. How much
energy does each parton in the dipole give to the emitted gluon? Obviously
the answer should be such that if the new gluon comes very close to one of the
partons, that parton should take most of the recoil, while the other one should
be less affected. There are, however, different ways to share the
amount of energy between the partons.

Consider the emission of gluon $n$ from the dipole $ij$. Let $r_{in}$ and
$r_{nj}$ denote the distance between the new gluon and parton $i$ and parton
$j$ respectively. Denote the recoil by $p_+''$, while $p_+'$ is the amount of
energy left after the emission, which implies that $p_+=p_+''+p_+'$. We then make the ansatz that 
\begin{equation}
p_{+i}''=\frac{r_{jn}}{r_{jn}+r_{in}}p_{+n} \; \; \textrm{and}\; \;
p_{+j}''=\frac{r_{in}}{r_{jn}+r_{in}}p_{+n}
\end{equation}

Thus we assume that $p_{+i}''/p_{+j}''=r_{jn}/r_{in}$. We have also studied alternative ways to share the recoils but the result does not depend sensitively on the exact formula. When we have
generated an emission we always check that $p_+''\leqslant p_+$, which
obviously must be satisfied.   

To make the analysis simple, transverse momentum, $\bold{p}_\perp$, is only approximately conserved in the program. Exact conservation of $\bold{p}_\perp$ would be possible but is not essential for the purpose of this study. When a new
gluon is emitted its transverse momentum is decided by
\begin{equation}
p_{\perp n}=\textrm{max}\biggl(\frac{1}{r_{in}},\frac{1}{r_{jn}}\biggr)
\end{equation}
   
The recoils on the emitters are given by 
\begin{eqnarray}
& &\textrm{If} \;  p_{\perp i}\leqslant \frac{1}{r_{in}} \textrm{then} \;
p_{\perp i}'=\frac{1}{r_{in}} \\
& &\textrm{If} \;  p_{\perp j}\leqslant \frac{1}{r_{jn}} \textrm{then} \;
p_{\perp j}'=\frac{1}{r_{jn}} \nonumber
\end{eqnarray}   
Otherwise we simply set $p_\perp =p_\perp'$, i.e we don't change the transverse
momenta. This simply means that the transverse momentum is decided only by the largest contribution. Hence
 the transverse momentum of a parton will always be determined by the shortest distance to
another parton, with which it has formed a dipole. Since we change $p_+$ and
$p_\perp$ we must, to be consistent, also change the rapidity $y$. This is
obviously done by setting 
\begin{equation}
y'=\textrm{ln}\frac{p_\perp'}{p_+'}\geqslant y
\end{equation}
The inequality is due to the fact that $p_\perp'\geqslant p_\perp$ and
$p_+'\leqslant p_+$. Therefore the rapidity will always increase due to a
recoil. There is however a problem with this; as we order the emissions in
rapidity the last emitted gluon will always have the largest rapidity, and we
would like it to remain so. Since the rapidity of the emitters
increase due to recoil, there is the possibility that one of the emitters, or
both of them, after the recoil ends up with a rapidity  larger than the
rapidity of the emitted gluon. The next time we emit a gluon, the rapidity
of that emitter would then determine the region in which the new emission can occur, in
a sense violating the ordering we had. To prevent this we require that
$y_i',y_j'\leqslant y_n$; if this constraint is not satisfied a new gluon is generated. Without this
constraint there would also be a lot of gluons in the final state which have
rapidities larger than the maximum rapidity, due to the recoils. 

\subsection{Upper Bound for the Dipole Size}

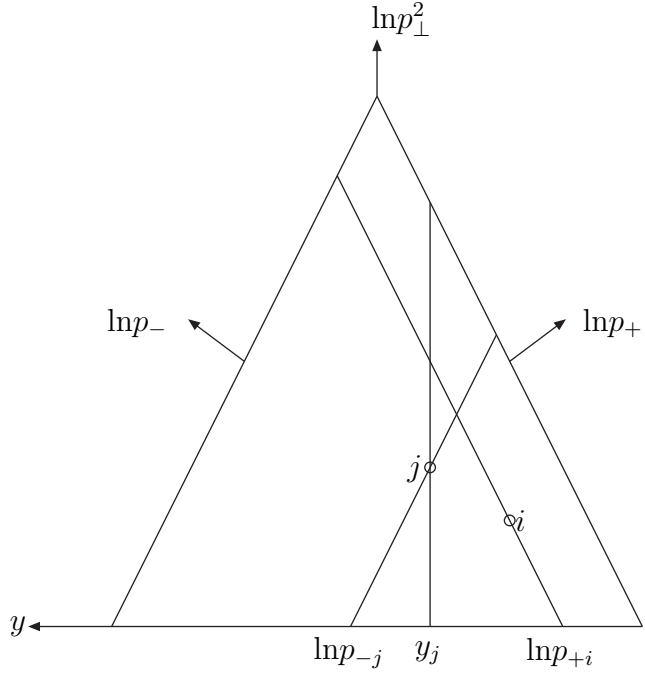
\begin{figure}[ty]
\begin{center}
\begin{picture}(400,260)(0,0)
\LongArrow(300,20)(70,20)
\Line(100,20)(200,220)
\Line(300,20)(200,220)
\LongArrow(200,220)(200,240)
\Text(210,250)[]{ln$p_\perp^2$}
\Text(65,20)[]{$y$}
\Text(290,135)[]{ln$p_+$}
\Text(110,135)[]{ln$p_-$}
\LongArrow(250,120)(270,135)
\LongArrow(150,120)(130,135)
\BCirc(250,60){2}
\BCirc(220,80){2}
\Line(190,20)(245,130)
\Line(220,180)(220,20)
\Line(270,20)(185,190)
\Text(255,60)[]{$i$}
\Text(215,80)[]{$j$}
\Text(220,10)[]{$y_j$}
\Text(270,10)[]{ln$p_{+i}$}
\Text(190,10)[]{ln$p_{-j}$}
\end{picture}
\end{center} 
\caption 
  {\emph{The triangular phase space for emissions. The
  two circles indicate the emitting partons, the vertical line through 
  circle j indicates the rapidity of $j$ while the on second line,
  $p_-$ is equal to $p_{-j}$ Emissions  between the two lines through $j$ are
  forbidden due to $p_-$ ordering, which limits the allowed rapidity interval. The line
  through i indicates its $p_+$ value.}} 
\label{pminus}
\end{figure}

The $p_+$ conservation, as we mentioned, will give a cutoff for small dipoles,
limiting the number of small dipoles. We can, however, also get
extremely large dipoles, especially when the avaliable energy is low and the
rapidity is not very large. This follows immediately from the fact that $\rho
(y)=e^{-y}/p_+$ gets very large for limited $y$ and small $p_+$. We can also take 
extremely small steps in rapidity because the phase space
avaliable permits this, see figure \ref{pminus}. When we are about to emit a new gluon we let the
computer generate an emission for each existing dipole, and then we choose the
emission which has the lowest $y$. Because of what we said above, the step
sizes in $y$ can, and indeed will, be extremely small which result in a huge
number of dipoles and very long running times. This can be fixed by imposing an
ordering in negative light cone-momentum, $p_-$, such that 
\begin{equation}
p_{-n}\geqslant max(p_{-i},p_{-j}) \label{eq:pmin}
\end{equation}

There is also a physical argument for doing this. As we described in section 3.1 we want the 
cascade to be right-left symmetric, and by imposing $p_-$ ordering we can achieve this.
This ordering will make very small steps in $y$ less probable, and it will also
limit the radius of a dipole from above. To see this let the larger negative
light-cone momenta be $p_-$ which implies 
\begin{equation}
p_{-n}\geqslant p_-\Rightarrow p_\perp e^y=\frac{1}{r}e^y\geqslant p_-\;
\textrm{hence}\; r\leqslant \frac{e^y}{p_-}
\end{equation}

We are limiting the small dipoles with the energy avaliable in the emitting
dipole, so the smallest dipole that can be created is the same for both 
partons of the mother dipole. A more realistic way might be to set
two different lower limits for the two partons. If one of them has
greater energy than the other, it should be possible for the emitted gluon to
form a smaller dipole with this parton than with the other one. Thus,
the energy of the gluon should be bounded by the energy of the two partons
separately, and not by their sum. We would then have
\begin{equation}
r_{in}\geqslant \frac{e^{-y}}{p_{+i}} \; \textrm{and} \; r_{jn}\geqslant
  \frac{e^{-y}}{p_{+j}}
\end{equation} 
This is however already satisfied in our program due to our constraint
$y_i'\leqslant y_n$, which is necessary for conserving the $y$ ordering. The constraint
\begin{eqnarray}
& &y_i'\leqslant y_n \; \; \textrm{implies} \; \; e^{-y_i'}\geqslant e^{-y_n}. \; \;
\textrm{The relation} \; \;
e^{-y_i'}=e^{-ln\frac{p_{\perp i}'}{p_{+i}'}}=\frac{p_{+i}'}{p_{\perp
    i}'} \; \textrm{then implies}\nonumber \\
& &\frac{1}{p_{\perp i}'}\geqslant
\frac{e^{-y_n}}{p_{+i}'}\geqslant \frac{e^{-y_n}}{p_{+i}}  \; \; \textrm{which together with the relation}\nonumber \\
& & \frac{1}{p_{\perp i}'}=\frac{1}{max\biggl(p_{\perp
    i},\frac{1}{r_{in}}\biggr)}\leqslant r_{in}  \\
& &\textrm{gives} \; r_{in}\geqslant \frac{e^{-y_n}}{p_{+i}} \nonumber 
\end{eqnarray} 

\subsection{Formulas for Generating y and r}

We have above described the main ideas of our program. In this
section we will write down the explicit equations used to generate $y$ and $\bold{r}$
values. As we mentioned in section 2.2, the $r_x$ and $r_y$ factors in~\eqref{eq:mueller2}
have the meaning of a splitting function. Taking into account also the Sudakov
factor, the total formula for emission is given by 
\begin{equation}
\frac{\bar{\alpha}}{2\pi}\int_{\rho (y'')}
d^2\bold{r}_n\frac{r_{ij}^2}{r_{in}^2r_{jn}^2}exp\biggl[-\frac{\bar{\alpha}}{2\pi}\int_{y'}^{y} dy''
\int_{\rho (y'')}d^2\bold{r}_n\frac{r_{ij}^2}{r_{in}^2r_{jn}^2} \biggr]
\label{eq:splittingweigth}
\end{equation}

The Sudakov factors in section 2.2 were trivial, since we had no $y$ dependence
on the $\bold{r}$ integral. Demanding energy conservation implies, however, that the lower limit of the integral
depends on $y$ and the Sudakov factor will therefore be more complex. To make
the calculations simpler, we set $\bold{r}_i=(0,0)$ and $\bold{r}_j=(1,0)$, so
that $r_{ij}=1$. This we can always do by applying a translation of
$-\bold{r}_i$, then rotating the vector $\bold{r'}_j=\bold{r'}_{ij}$ onto the
$x$-axis and finally by scaling with $r_{ij}$. After we have generated 
values for $\bold{r}$ we transform the system back to the original one. To
generate the $y$ and $\bold{r}$ values we use the well known veto algorithm,
and the $y$ values are given by the distribution
\begin{equation}
y=-\textrm{ln}\sqrt{1+\frac{p_+^2}{4}}+\sqrt{(\textrm{ln}\sqrt{1+\frac{p_+^2}{4}}+y')^2-\frac{\textrm{ln}R}{4\bar{\alpha}}}\label{eq:rap}
\end{equation}
where $R$ is a random number, $y'$ is the maximum rapidity before the
emission, and $p_+$ is the total energy of the emitting dipole. The values for $r_x$ and $r_y$ are chosen from the distribution
\begin{eqnarray}
& &r_x=\sqrt{\frac{\rho^2(1+\frac{1}{4\rho^2})^{R_2}}{1+4\rho^2-4\rho^2(1+\frac{1}{4\rho^2})^{R_2}}}\cos{2\pi
  R_1} \label{eq:r}\\
& &r_y=\sqrt{\frac{\rho^2(1+\frac{1}{4\rho^2})^{R_2}}{1+4\rho^2-4\rho^2(1+\frac{1}{4\rho^2})^{R_2}}}\sin{2\pi
  R_1} \nonumber
\end{eqnarray}
where $\rho =\rho (y)$ and $R_1$ and $R_2$ are two, independent, random
numbers. Note that we use the generated $y$-value for generating $r_x$ and
$r_y$. Finally, we accept the generated values with the 
probability
\begin{equation}
\frac{\textrm{ln}(1+\frac{p_+^2}{4}e^{2y})((r_x-1)^2+r_y^2+0.25)(r_x^2+r_y^2+0.25)}{(\textrm{ln}(1+\frac{p_+^2}{4})+2y)\biggl(((r_x-1)^2+r_y^2)((r_x-1)^2+r_y^2+0.25)+(r_x^2+r_y^2+0.25)(r_x^2+r_y^2)\biggr)}\label{eq:sann}
\end{equation}
This formula is derived in the appendix.

\subsection{Summary of the Program}

In this section we will summarize the important points from the earlier
sections about the MC program. The algorithm is

\begin{enumerate}
\item 
For each dipole, pick a $y$ value from ~\eqref{eq:rap} and $r_x$ and $r_y$ from ~\eqref{eq:r}.
\item
For these values calculate the weight in~\eqref{eq:sann}. If ~\eqref{eq:sann} $<R$ go
back to 1 and start over. Otherwise go to 3.
\item
If $p_{+i}\leqslant p_{+i}''$ or $p_{+j}\leqslant p_{+j}''$ go back to 1 and
start over. Otherwise go to 4.
\item
Check ~\eqref{eq:pmin}. If the inequality is not satisfied, then go back to 1 and start
over. Otherwise go to 5.
\item
Check if $y_i',y_j'\leqslant y$. If the inequality is not satisfied, then go back to 1
and start over. Otherwise generate a gluon.
\item
Set the new values for $p_{+i,j}$, $p_{\perp i,j}$ and $y_{i,j}$. 
\item
Select the dipole which gives the lowest $y$.
\item
If $y < y_{max}$ let the selected dipole emit the generated gluon. Otherwise stop the process.
\item
Delete all other generated gluons and continue.
\end{enumerate}

\section{Applications of the MC Program}

In this section we consider the applications of the program. We use the
program to calculate cross-sections for two processes, onium-nucleus
scattering and onium-onium scattering. 

\subsection{Onium-Onium Scattering}

 We start with onium-onium collisions. There has been other
studies of this process, in particular there is another MC program called OEDIPUS,
developed by Salam, \cite{e8}, which is also based on Mueller's Dipole formulation,
but without energy conservation. To study this process we 
generate two onium states independent of each other, one developed up to a rapidity $y$ and the other
developed up to $Y - y$, and after that we let them collide. The collisions will
occur between the dipoles from the two onium states and to calculate the cross
section we need an expression for the $S$ matrix. It is here in order to present the 
theoretical background for how we arrive at an expression for the S matrix. The amplitude for one pomeron exchange between two colliding onia is, in the BFKL approximation, given by
\begin{equation}
F^{(1)}(\bold{r}_1,\bold{r}_2,\bold{b},Y)=-\int \frac{d^2 \bold{c} d^2\bold{c}'}{2\pi c^2 2\pi c'^2}d^2 \bold{r}d^2\bold{r}'
f(\bold{r}-\bold{r}',\bold{c},\bold{c}')n(c,r_1,r,y)n(c',r_2,\vert \bold{r}'-\bold{b}\vert,Y-y)
\end{equation}
Here $r_1$ and $r_2$ are the sizes for the two initial dipoles, $c$ is the size of a dipole in the right moving onium while $c'$ is the size of a dipole in the left moving onium and $\bold{r}$ and $\bold{r}'$ are their positions relative the initial dipoles respectively. $n(c,r_1,r,y)$ is the density of dipoles of size $c$ and at a distance $r$ from the initial dipole after evolution through some rapidity $y$. $Y$ is the maximum rapidity determined by the cms energy. The expression for the dipole-dipole interaction is given in \cite{e3,e8,e9}, where it is written as
the square of the two-dimensional dipole-dipole potential
\begin{equation}
f(\bold{r},\bold{c},\bold{c}')=\frac{\alpha_s^2}{2}\biggl[\log
\frac{\vert \bold{r} + \bold{c}/2 -\bold{c}'/2 \vert \vert \bold{r} - \bold{c}/2
  + \bold{c}'/2\vert}{\vert \bold{r} + \bold{c}/2 + \bold{c}'/2 \vert \vert \bold{r} - \bold{c}/2
  - \bold{c}'/2\vert}\biggr]^2
\label{eq:salam}
\end{equation}   

Onium-onium scattering can be formally expressed in the operator formalism presented in \cite{e3}. To that end we define the operators $\mathcal{A}^\dagger (\bold{r},\bold{b})$ and $\mathcal{A}(\bold{r},\bold{b})$, where the first one creates a dipole of size $r$ and with impact parameter $\bold{b}$ while the second one destroys the same dipole. One usually uses $\mathcal{A}^\dagger$ and $\mathcal{A}$ for the right moving onium and $\mathcal{D}^\dagger$ and $\mathcal{D}$ for the left moving onium. In the operator formalism there are two vertices, the first one, $\mathcal{V}_1$, is the vertex for a dipole splitting process while the second one, $\mathcal{V}_2$ is the vertex containing the virtual corrections. These are given by
\begin{eqnarray}
\mathcal{V}_1&=&\bar{\bar{\alpha}}\int d^2\bold{r}_{ij}d^2\bold{r}_{in}d^2\bold{r}_{jn}d^2\bold{b}_{ij}\delta (\bold{r}_{ij}+\bold{r}_{in}+\bold{r}_{jn})
\frac{r_{ij}^2}{r_{in}^2r_{jn}^2}\times \nonumber \\
& &\times \mathcal{A}^\dagger (\bold{r}_{in},\bold{b}_{ij}-\frac{1}{2}\bold{r}_{jn})\mathcal{A}^\dagger (\bold{r}_{jn},\bold{b}_{ij}+\frac{1}{2}\bold{r}_{in})
\mathcal{A}(\bold{r}_{ij},\bold{b}_{ij})
\label{eq:op1}
\end{eqnarray}
and 
\begin{equation}
\mathcal{V}_2=-2\bar{\alpha}\int d^2 \bold{r}_{ij}d^2\bold{b}_{ij}\textrm{ln}\biggl(\frac{r_{ij}}{\rho}\biggr)\mathcal{A}^\dagger (\bold{r}_{ij}, \bold{b}_{ij})\mathcal{A}(\bold{r}_{ij}, \bold{b}_{ij})
\label{eq:op2}
\end{equation}
where, in~\eqref{eq:op1}, we have defined $\bar{\bar{\alpha}}=\bar{\alpha}/2\pi$.

Equation~\eqref{eq:op1} is just the vertex for the splitting of dipole $ij$ into $in$ and $jn$, and here the delta function guarantees that the dipoles are connected. The second vertex simply takes care of the disappearing mother dipole, which is indicadet by the minus sign. This must be taken into account just as we did in section 2.1.1. We define $\mathcal{V}_R=\mathcal{V}_1+\mathcal{V}_2$ for the right moving onium and $\mathcal{V}_L$ is defined similarly. The amplitude for multiple scatterings is then given by
\begin{equation}
F^{(k)}(\bold{b},Y)=\langle 0\vert e^{\mathcal{A}_1+\mathcal{D}_1}\frac{(-f)^k}{k!}e^{y\mathcal{V}_R+(Y-y)\mathcal{V}_L}\mathcal{D}^\dagger (\bold{b},\bold{r}_2)
\mathcal{A}^\dagger (\bold{0},\bold{r}_1)\vert 0\rangle
\label{eq:ampk}
\end{equation} 
where we always choose our coordinate system such that $r_1$ has impact parameter $\bold{0}$. Here $\vert 0\rangle$ is the state defined by $\mathcal{A}\vert 0\rangle=\mathcal{D}\vert 0\rangle =0$. $\mathcal{A}_1$ is defined by
\begin{equation}
\mathcal{A}_1=\int d^2\bold{b}d^2\bold{c}\mathcal{A}(\bold{b},\bold{c})
\end{equation}
with a similar expression for $\mathcal{D}_1$. In~\eqref{eq:ampk} the factor $f$ describes the interaction, the factor involving $\mathcal{V}$ describes the evolution of the onia states where we create any given number of dipoles for all possible positions and sizes. $\mathcal{D}^\dagger$ and $\mathcal{A}^\dagger$  simply create the initial dipoles while $e^{\mathcal{A}_1+\mathcal{D}_1}$ searches for any given number of dipoles for all possible positions and sizes. The quantity $f$, in~\eqref{eq:ampk}, is given by
\begin{equation}
f=\int d^2\bold{r}d^2\bold{r}'d^2\bold{c}d^2\bold{c}'f(\bold{r}-\bold{r}',\bold{c},\bold{c}')\mathcal{D}^\dagger (\bold{r}',\bold{c}')\mathcal{D}(\bold{r}',\bold{c}')\mathcal{A}^\dagger (\bold{r},\bold{c})\mathcal{A}(\bold{r},\bold{c})
\end{equation}
The meaning of the integral is that we count the number of dipoles for all possible positions and sizes in the right and left moving onia ($\mathcal{D}^\dagger\mathcal{D}$ and $\mathcal{A}^\dagger\mathcal{A}$ are approximately the number operators). We then multiply by the interaction amplitude for the dipoles and finally we sum all contributions. Thus $f$ has the meaning of a total amplitude. Summing pomeron exchanges to all orders one arrives at the formula for the $S$ matrix (which is our goal)
\begin{eqnarray}
S(\bold{r}_1,\bold{r}_2, \bold{b},Y)&=& 1+F(\bold{r}_1,\bold{r}_2, \bold{b},Y)\nonumber \\
&=&\langle 0\vert e^{\mathcal{A}_1+\mathcal{D}_1}e^{-f}e^{y\mathcal{V}_R+(Y-y)\mathcal{V}_L}\mathcal{D}^\dagger (\bold{b},\bold{r}_2)
\mathcal{A}^\dagger (\bold{0},\bold{r}_1)\vert 0\rangle
\end{eqnarray}

To be used in a MC simulation however, this expression for the $S$ matrix is not so good. To find a more suitable expression we first note that the probability to find $n$ dipoles in an onium state with initial dipole given by $b_0$ and $r_0$, and at a rapidity $Y$, is in the operator formalism given by
\begin{equation}
P_n(\bold{b}_0,\bold{r}_0,Y)=\langle n\vert e^{Y\mathcal{V}}\mathcal{A}^\dagger(\bold{b}_0,\bold{r}_0)\vert 0\rangle
\end{equation}

Assume that we connect the two onium states at the center, i.e $y=Y/2$. Then the $S$ matrix can be written in the following form 
\begin{equation}
S(\bold{r}_1,\bold{r}_2, \bold{b},Y)=\sum_{a,a'}P_a(\bold{r},\bold{0},Y/2)P_{a'}(\bold{r}',\bold{b}',Y/2)e^{-f_{a,a'}}
\label{eq:Smatrix}
\end{equation}
 where the sum runs over all possible configurations, $a$ and $a'$, and $f_{a,a'}$ is given by
\begin{equation}
f_{a,a'} =
\sum_{i=1}^{N_a}\sum_{j=1}^{N_{a'}}f(\bold{r}_i-\bold{r}'_j,\bold{c}_i,\bold{c}'_j)
\end{equation}
$N_a$($N_{a'}$) is the total number of dipoles for the configuration $a$($a'$). In arriving at the formula for the $S$ matrix, it is important that one sums the contributions from the multiple scatterings for a particular configuration before the averaging over all possible configurations is done. The reason to this is that the multiple scattering series does not converge, which is due to the fact that rare configurations give large contributions to the series, especially for higher orders. It is easy to see that the $S$ matrix, as given in~\eqref{eq:Smatrix}, satisfies the unitarity constraints
\begin{equation}
 S\leqslant \sum_{a,a'} P_aP_{a'} \leqslant 1
\end{equation}
since $f_{a,a'}\geqslant 0$. The unitarised amplitude is thus given by
\begin{equation}
\vert F\vert=1-S = 1 - \sum_{a,a'}P_aP_{a'}e^{-f_{a,a'}}= \sum_{a,a'}P_aP_{a'}(1-e^{-f_{a,a'}})
\label{eq:uniamp}
\end{equation}

A multiple scattering series for the amplitude emerges when one expands the $e^{-f}$ term in the $S$ matrix. Therefore the one pomeron amplitude is given by
\begin{equation}
\vert F^{(1)}\vert= 1 - \sum_{a,a'}P_aP_{a'}(1-f_{a,a'})=\sum_{a,a'}P_aP_{a'}(1-(1-f_{a,a'}))=\sum_{a,a'}P_aP_{a'}f_{a,a'}
\label{eq:unione}
\end{equation}
The factor $1-e^{-f_{a,a'}}$, in~\eqref{eq:uniamp}, has the meaning of the total reaction probability for the onium-onium collision.
Therefore all we need to do is to calculate $1-e^{-f_{a,a'}}$ for randomly chosen initial conditions, and the unitarised cross-section for
onium-onium scattering will be given by 
\begin{equation}
\sigma = \int d^2\bold{b}(1-e^{-f_{a,a'}}) 
\label{eq:sigmaonioni}
\end{equation}
To evaluate the integral one must choose a finite area over which the random positions are chosen. 
This is quite easy to do, since the amplitude decreases very fast
for large $b$. By checking the amplitude for large $b$ one easily finds a good value for the
maximal value of $b$. In order to obtain the one pomeron amplitude we just do the same thing, but with $1-e^{-f_{a,a'}}$ replaced by $f_{a,a'}$, which is easily seen from~\eqref{eq:unione}. Therefore the one pomeron cross section is given by  
\begin{equation}
\sigma^{(1)} = \int d^2\bold{b}f_{a,a'}
\end{equation}

Some comments on formula
~\eqref{eq:salam} are in order here. The four terms in the logarithm are the
four distances between the ends of the dipoles, see figure 7. 
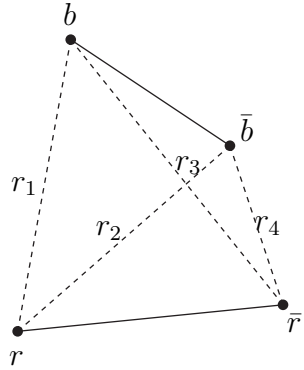
\begin{figure}
\begin{center}
\begin{picture}(400,150)(0,0)
\Line(100,20)(200,30)
\Line(120,130)(180,90)
\Text(103,75)[]{$r_1$}
\Text(135,58)[]{$r_2$}
\Text(165,83)[]{$r_3$}
\Text(195,60)[]{$r_4$}
\Text(100,10)[]{$r$}
\Text(205,23)[]{$\bar{r}$}
\Text(120,140)[]{$b$}
\Text(187,97)[]{$\bar{b}$}
\DashLine(100,20)(120,130){2}
\DashLine(100,20)(180,90){2}
\DashLine(120,130)(200,30){2}
\DashLine(180,90)(200,30){2}
\Vertex(100,20){2}
\Vertex(120,130){2}
\Vertex(180,90){2}
\Vertex(200,30){2}
\end{picture}
\end{center} 
\caption 
  {\emph{With the notations in the figure the logarithm term in the interaction is
  $\log \frac{r_2r_3}{r_1r_4}$.}} 
\end{figure}
\begin{figure}
\begin{center}
    \includegraphics[angle=0, scale=1.28]{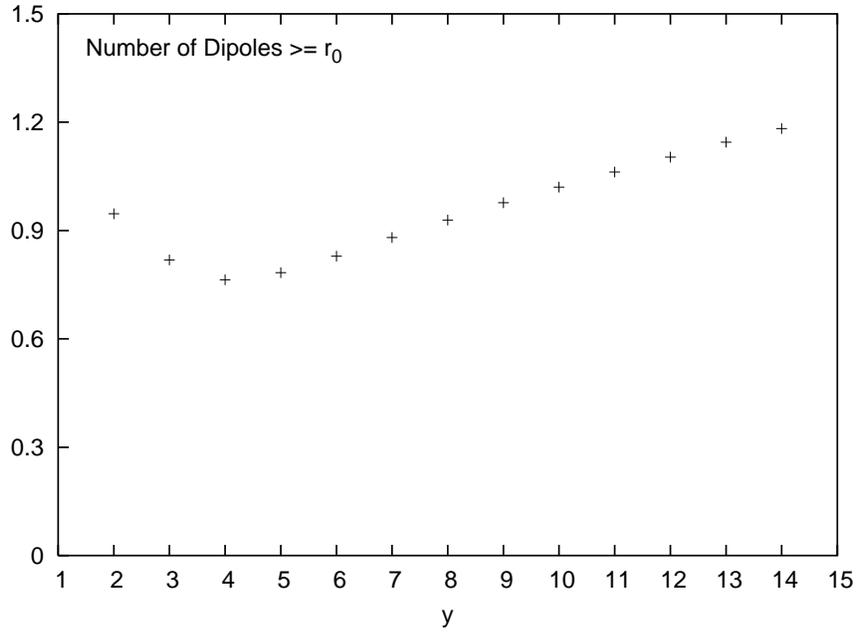}
\end{center}
    \caption{\emph{The average number of dipoles with sizes greater than or equal to the initial dipole.
 Although the number of dipoles increase exponentially, the number of large dipoles increase very slowly.}}
\label{numberdip}
\end{figure} 
If the size of one of the dipoles is much smaller than the other $r_1\rightarrow r_2$
and $r_4\rightarrow r_3$ and thus the amplitude will tend to zero. Therefore
small dipoles interact weakly. Also, if the size of one of the dipoles tends
to infinity and one of its ends is not very close to the other dipole, the
distances will again be mutually equal and the interaction very weak. 

\begin{figure}
\begin{center}
    \includegraphics[angle=0, scale=1.28]{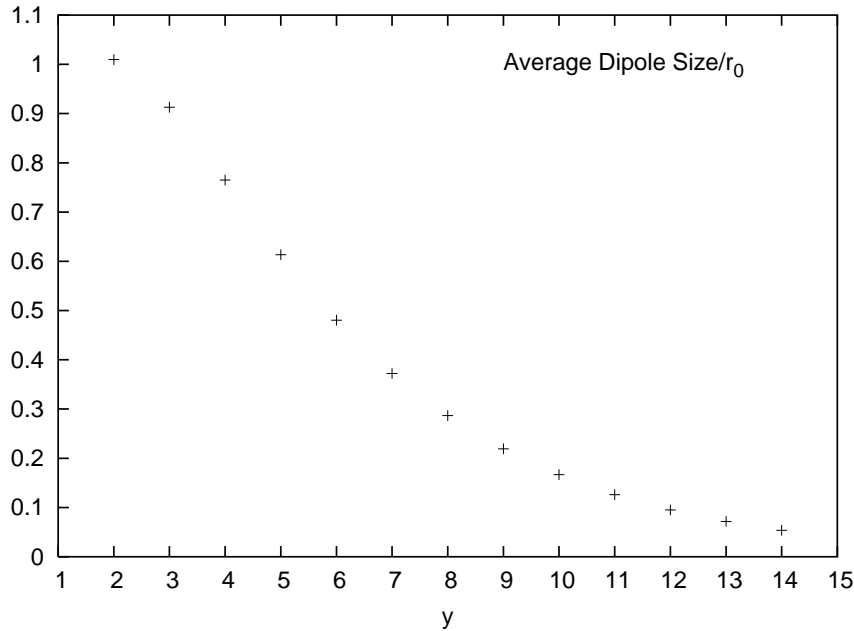}
\end{center}
    \caption{\emph{The average dipole size divided by the initial dipole size  as a function of the rapidity, $y$. We see that the dipole size drops rapidly as $y$ increases, just as we expect.}}
\label{dipolesize}
\end{figure} 

The number of dipoles in an onium state increases exponentially with
$y$. However, as $y$ increases it is easy to see that the average dipole size
will decrease very fast. This is so because $\rho=e^{-y}/p_+$ gets very small for large $y$, 
and the probability to get small dipoles increase as $\rho$ decrease. Since $\rho$ drops exponentially we also expect
the average dipole size to drop rapidly. This can be seen in figure \ref{dipolesize}. As the number of
dipoles increases, the interaction term between the two onium states will of
course increase, and since the number of dipoles increases exponentially one would
expect that the cross section also increases exponentially with $y$. However,
because of what we just said above, and the fact that interactions
involving small dipoles are very weak, the growth rate is reduced. In fact, the number of large dipoles rises very slowly (with large we here
mean dipoles which are larger than the initial dipole), see figure 
\ref{numberdip}.  
\begin{figure}
\begin{center}
    \includegraphics[angle=0, scale=1.28]{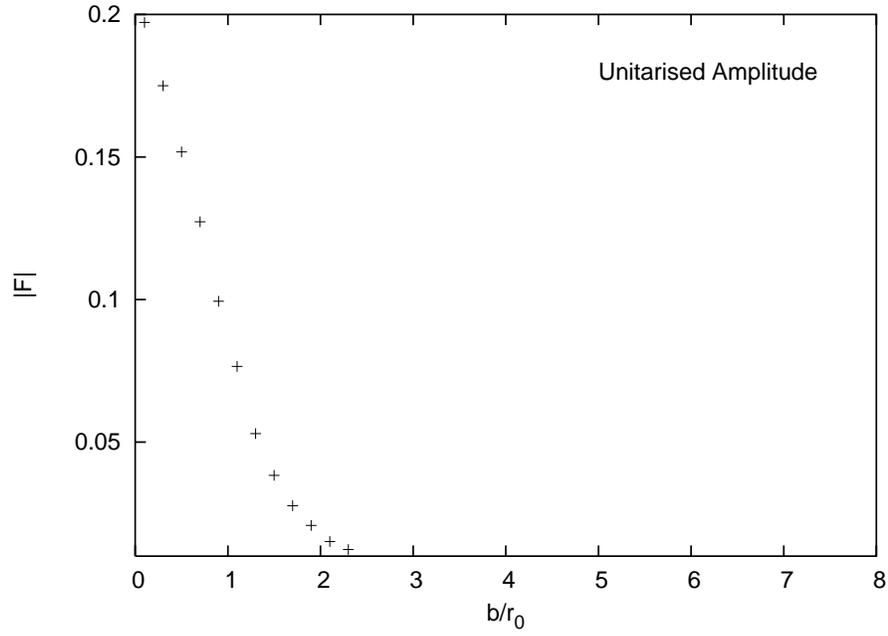}
\end{center}
    \caption{\emph{The unitarised amplitude as a function of $b_0/r_0$, $r_0$ being the size of the two initial dipoles, for Y=14.}}
  \label{uniamp}
\end{figure} 
In figure \ref{uniamp} we plot the unitarised amplitude as a function of the impact parameter $b_0$, at a rapidity $Y=14$, and for $r_1=r_2=r_0$. We see that the amplitude is well below 1 for all impact parameters and that it drops quite fast with increasing $b_0$. We also plot the one pomeron amplitude as a function of $b_0$ and for  $Y=14$, see figure \ref{1pomamp}. It is interesting to see that the one pomeron amplitude is below 1 for all $b_0$ and that it is very similar to the unitarised amplitude.  
\begin{figure}
\begin{center}
    \includegraphics[angle=0, scale=1.28]{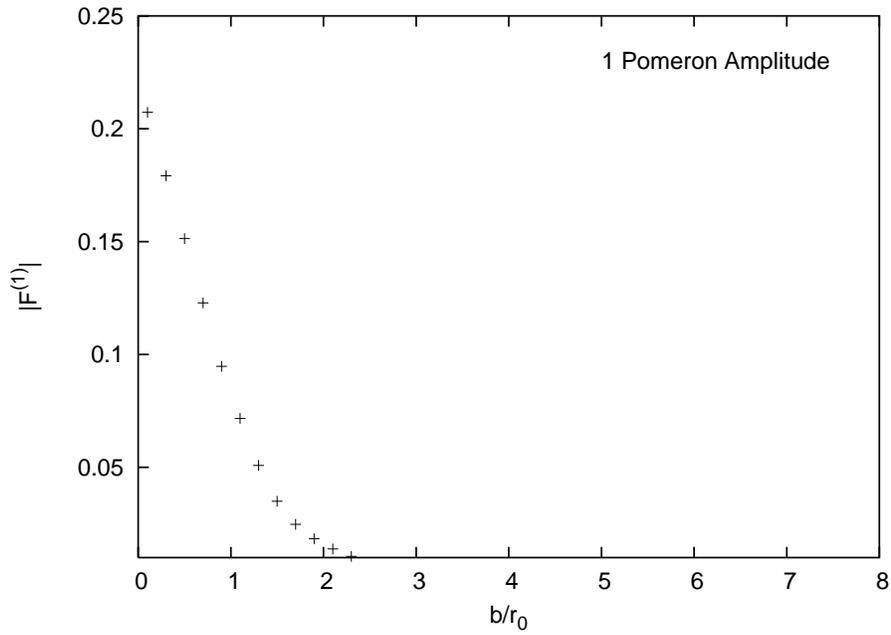}
\end{center}
    \caption{\emph{The One Pomeron amplitude as a function of $b_0/r_0$ for Y=14.}}
  \label{1pomamp}
\end{figure} 

This is in sharp contrast to previous studies, for example \cite{e9}, in which energy conservation is not included. Here the difference between the unitarised and the one pomeron amplitudes is quite large. In \cite{e9} one can also see that the one pomeron amplitude is above 1 for $b_0\lesssim r_0$ and that both amplitudes are above zero for $b_0\lesssim 10r_0$.  

\subsection{Onium-Nucleus Scattering}

As a second application we study scattering off a nucleus. As in section 2 we consider a nucleus
of cylinder shape with some radius $B$. In the transverse plane the nucleus
will look like a circle with radius $B$. The center of the nucleus is located
at $\bold{b}_0$. The position of our initial dipole is kept fixed, its
coordinates are given by $\bold{r}_0=(-0.5r_{10},0)$ and $\bold{r}_1=(0.5r_{10},0)$. The
impact parameter is then simply equal to $\bold{b}_0$. In order to calculate
the cross section for this event we have to find an expression for the
interaction between the dipoles and the nucleus. First of all one should note
that the amplitude, for dipole-nucleus scattering, is going to depend on five
parameters. Two of the parameters give the position of the center of the
nucleus in the transverse space, namely $b_0 \equiv \vert \bold{b}_0 \vert$
and $\psi_b$, the third parameter is the size of the initial dipole,
$r_0$, the fourth is the angle describing the relative orientation between the
target and the initial dipole and finally the fifth parameter is the rapidity. The relative
orientation however, is not needed when the nucleus is cylindrical, which is
quite obvious. The rapidity and the size of the initial dipole come
in naturally in a MC simulation and are not 
needed explicitly in the interaction formulas. Therefore we are left with
two parameters instead of five. 

As mentioned above we picture the nucleus as a circle in transverse space. We will view the nucleus
as a collection of color dipoles and the interaction between an incoming dipole and the nucleus
will be given by the interaction between the incoming dipole and the target dipole in the nucleus,
convoluted with the dipole distribution of the nucleus. The interaction formula between two dipoles
is again given by~\eqref{eq:salam}. The dipole distribution can be chosen in a variety of ways but
 not arbitrarily. First of all, we want the dipole density to go to zero as the dipole
size goes to zero. This is because, as mentioned in section 3.1, small dipoles correspond to hard gluons and if we allow too small dipoles in the nucleus we will violate energy conservation. The dipole density should also go to zero when the dipole size approaches
infinity, we do not want dipoles which are larger than the nucleus. The dipole density should
also be large at the center of the nucleus, an incoming dipole with zero impact parameter should
have a larger chance to interact with the nucleus than a dipole with non zero impact parameter.
Let $\bold{b}_i$ be the impact parameter of the target dipole $i$ , i.e a dipole belonging the nucleus.
The position of dipole $i$ relative to the center of the nucleus is then simply given by
$\bold{b}=\bold{b}_i-\bold{b}_0$. We also expect the dipole density to drop quite fast for large $b$.
With these considerations we have chosen the following distribution 
\begin{equation}
d\mathcal{P}=d^2r_id^2br_i^2e^{-r_i^4}e^{-\frac{2b^2}{B^2}}
\end{equation}

Note that this whole expression is dimensionless, the $r_i$ and $b$ are all divided by a unit parameter
with the dimension length. Of course, in this expression, there should also be a normalization constant, 
which may come from energy conservation. We are here less interested in the precise value of the amplitude however, primarily we want to investigate how the amplitude behaves. In future studies a normalization constant should be included. In our calculations we have studied an example where $B$ is equal to five unit lengths. We see that the 
nucleus will consist of a collection of dipoles, with sizes small compared to the nucleus,
and a higher density near the center. The interaction between an incoming dipole $j$ and the nucleus is then given by
\begin{equation}
\bar{f}(\bold{b}_j-\bold{b}-\bold{b}_0,\bold{r}_i,\bold{r}_j)=\int d\mathcal{P}f(\bold{b}_j-\bold{b}-\bold{b}_0,\bold{r}_i,\bold{r}_j)
\label{eq:nucl1}
\end{equation}

Once we have this expression we just sum over all incoming dipoles $j$ and the total cross section is,
in analogy to~\eqref{eq:sigmaonioni}, given by 
\begin{equation}
\sigma=\int d^2b_0\biggl[1-exp\biggl(-\sum_{j}\bar{f}(\bold{b}_j-\bold{b}-\bold{b}_0,\bold{r}_i,\bold{r}_j)\biggr)\biggr]
\label{eq:nucl2}
\end{equation}         
The amplitude contains multiple pomeron exchanges up to all orders and it satisfies the unitarity bounds. We calculate~\eqref{eq:nucl1} 
and~\eqref{eq:nucl2} using MC techniques. For this one has to choose a finite area in which the last integral is evaluated.
This is quite easy to do since the amplitude falls off quite fast for large impact parameters. It is usually more than enough to take $b_{0max}\approx5(r_0+B)$.  

\subsection{Results}

In this section we present the results we obtained from our program. To start we like to point out that all calculations have been performed using a fixed coupling constant, $\bar{\alpha}=0.2$. Although a running coupling constant would obviously be more justified, we have avoided it in order to make the whole Monte Carlo program simpler. As we described in the last section, the main objective has been to obtain the cross-sections for the events we studied.
Therefore we start by presenting the cross-section for onium-onium scattering for different configurations. 

\begin{figure}
\begin{center}
  \includegraphics[angle=270, scale=0.7]{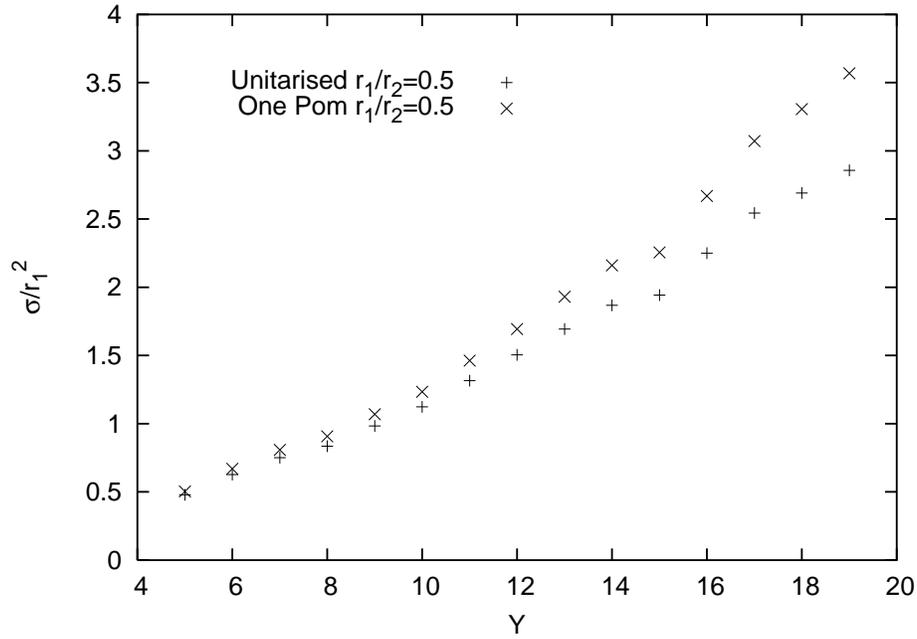}
\end{center}
\caption{\emph{Comparing the cross-section calculated from one pomeron and unitarised amplitudes, as a function of the rapidity $Y$.}}
\label{fig:oo1}
\end{figure}
  
In figure \ref{fig:oo1} we compare cross-sections obtained from one-pomeron and unitarised amplitudes.
We see that the difference, hence the saturation, is quite small. We could observe this already in 
section 4.1, where we compared the amplitudes as functions of the impact parameter. Observe that the scale in figure \ref{fig:oo1} is not logarithmic like in other figures. If plotted on logarithmic scale the difference between the cross-sections would barely be visible. One can see that the difference increases with $Y$, this is no surprise since $f$ will increase with $Y$ and therefore the difference between $f$ and $(1 - exp(-f))$ will also increase. 

\begin{figure}
\begin{center}
  \includegraphics[angle=270, scale=0.7]{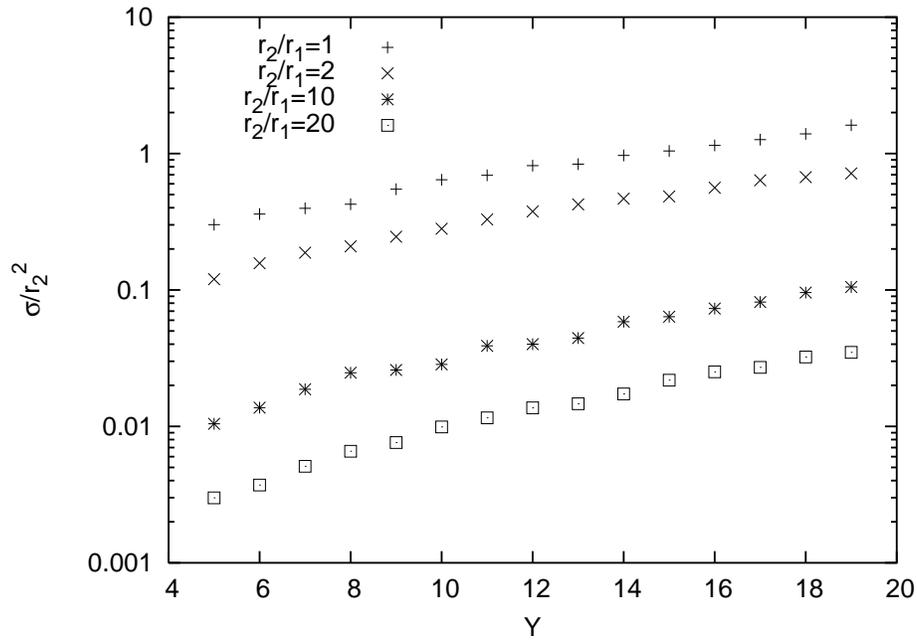}
\end{center}
\caption{\emph{$\sigma/r_2^2$ for various initial conditions as function of the rapidity Y. Here $r_1$ is kept fixed while we vary $r_2$. All calculations are done using the unitarised amplitude and for $y=Y/2$.}}
\label{fig:oo2}
\end{figure} 
Figure \ref{fig:oo2} shows the cross-sections for different initial conditions obtained from our MC program. The difference
compared to previous studies, \cite{e9}, is clear, our cross-sections grow much more slowly. One of the reasons for this is, as explained in section 4.1, that the average dipole size decrease rapidly as $Y$ increases and since small dipoles interact weakly, the growth of the cross section is reduced. The other reason is discussed below.

\begin{figure}
\begin{center}
  \includegraphics[angle=270, scale=0.7]{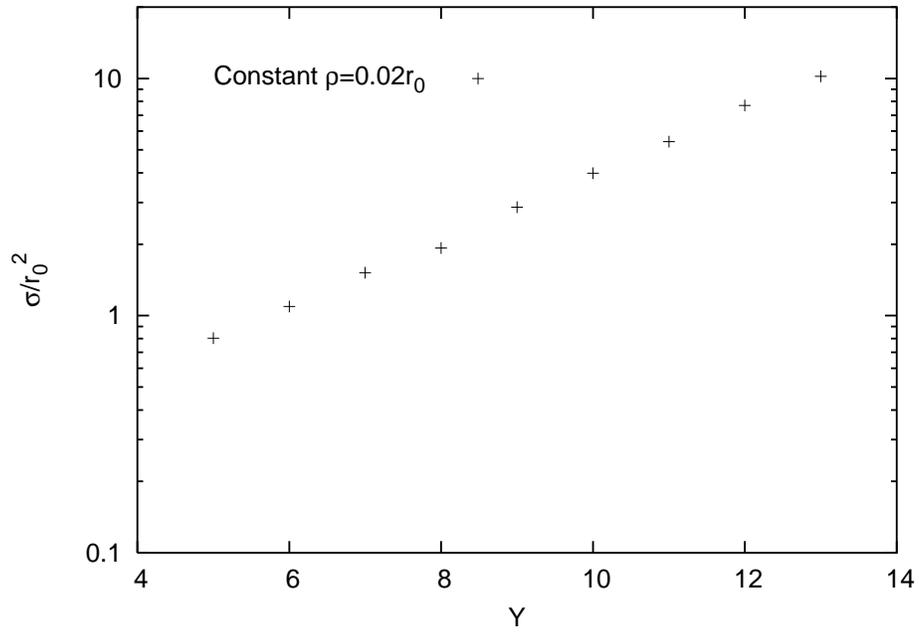}
\end{center}
\caption{\emph{$\sigma/r_0^2$, $r_0$ being the size of the two initial dipoles, using the unitarised amplitude, as a function of rapidity Y and without energy conservation.Note that the cross-section grows more rapidly compared to figure 13. }}
\label{fig:pureoo}
\end{figure}

In figure \ref{fig:pureoo} we show $\sigma/r^2$ as a function of the rapidity calculated for 
onium evolution with constant ultraviolet cut-off $\rho$, i.e without energy conservation.
Comparing with figure \ref{fig:oo2} we clearly see that $\sigma$ grows much faster in this case.

It is again interesting to see that the unitarised cross-section in figure \ref{fig:pureoo} grows even faster than the one pomeron cross-sections in figure \ref{fig:oo1}, we could see a hint of this already in section 4 where we plotted the amplitudes. We see that energy conservation has a non-negligible effect on the growth of the cross section. It seem as if one does not even has to use the unitarised amplitude, at least for the $Y$ intervals we investigated.   
\begin{figure}
\begin{center}
    \includegraphics[angle=0, scale=1.28]{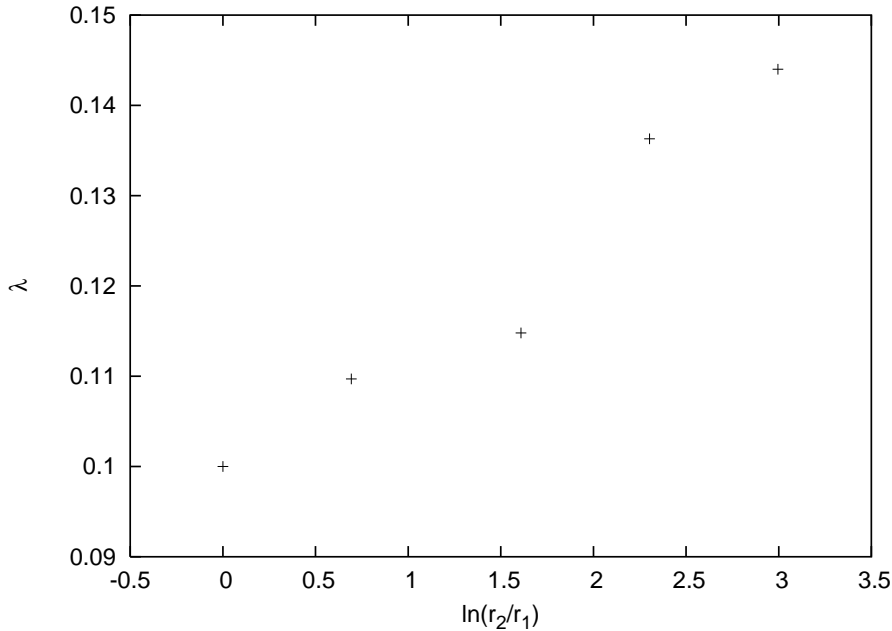}
\end{center}
    \caption{\emph{$\lambda$, plotted as a function of $r_2/r_1$, calculated from the unitarised cross-section for onium-onium collisions.}}
  \label{fig:lambda}
\end{figure}

Recall from section 2.1.2 that the gluon distribution function in the BFKL region is given by $G\sim x^{-\lambda}$. Here ln$1/x\sim y$, which gives $G\sim e^{\lambda y}$. Therefore the BFKL equation predicts an exponential rise of the cross section, $\sigma \sim e^{\lambda y}$. Thus ln$\sigma \sim \lambda y$ where, in the leading log approximation, $\lambda$ is given by $\frac{4\alpha_sC_A}{\pi}$ln2$=4\bar{\alpha}$ln2$\approx 0.55$ for $\bar{\alpha}=0.2$. By trying to fit a straight line for ln$\sigma$ between $Y=10$ and $Y=19$ we have obtained values for $\lambda$. The results are shown in figure \ref{fig:lambda}. We see that the $\lambda$ we have obtained is much lower than $\lambda_{BFKL}$. We explained one of the reasons for this above. The second reason is that  the steps in $y$, $\Delta y$, gets larger when we have energy conservation, which results in fewer dipoles. That $\Delta y$ gets larger can be seen from the phase space diagram in figure \ref{pminus}. Since we have both $p_+$ and $p_-$ conservation the avaliable phase space for emitting a gluon with a small $\Delta y$ is small. Therefore it is more likely to take bigger steps in $y$. If $\Delta y$ is the typical rapidity interval where the number of dipoles double, we can estimate the total number of dipoles with $N \sim 2^{Y/\Delta y}$. This can be written as $N \sim exp\biggl(\frac{\textrm{ln}2}{\Delta y}Y\biggr)$. Therefore we get $\lambda \sim$ ln$2/\Delta y$. It is not really true that $\sigma$ increase proportional to $N$ as the average dipole size is also important. The large $\Delta y$ does, however, give a very essential contribution to the reduction of the parameter $\lambda$. Thus it is expected that we obtain a lower $\lambda$ than in \cite{e9}. This is also expected from theoretical considerations. Taking into account higher order corrections reduces the value of $\lambda_{BFKL}$. For values, $\alpha_S \approx 0.2$, the value for $\lambda$ in NLO is even negative. This is of course not so realistic and there are different calculations that give some other values but it gives an idea of how $\lambda$ behaves. A large fraction of these higher order corrections are related to energy conservation, and therefore it is expected that one obtains lower values for $\lambda$. Observe that our values for $\lambda$ are not exact, there is some uncertainty in trying to fit a straight line for ln$\sigma$. We have calculated $\lambda$ also for constant $\rho$, and, for $\rho =0.02r_0$, $r_0$ being the size of the two initial dipoles, we get $\lambda=0.32$. This is a bit lower than $\lambda_{BFKL}=0.55$ but the value of $\lambda_{BFKL}$ is valid in the limit $\rho \rightarrow 0$.     
 
\begin{figure}
\begin{center}
  \includegraphics[angle=270, scale=0.7]{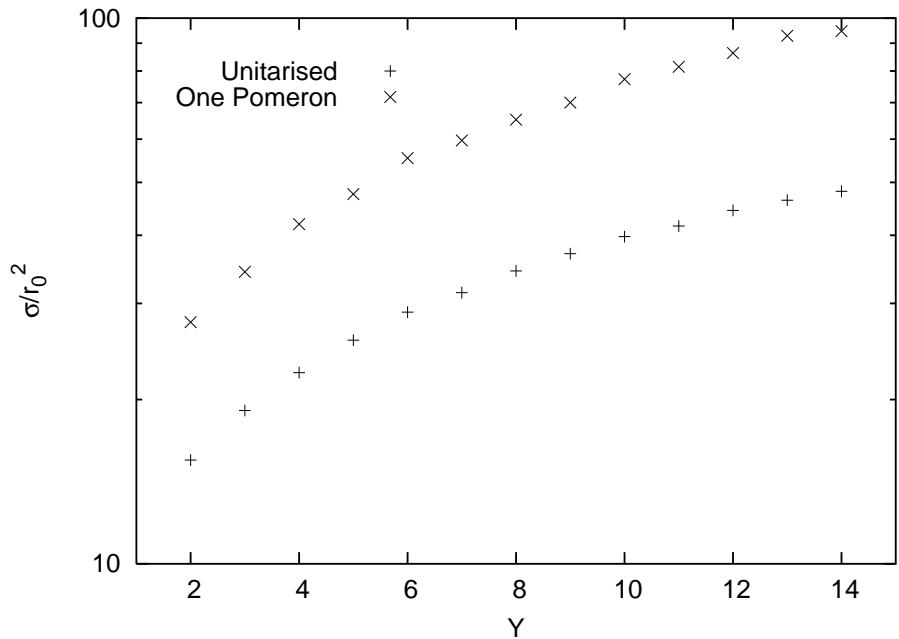}
\end{center}
\caption{\emph{$\sigma/r_0^2$ for unitarised and one pomeron calculations for onium-nucleus scattering as
function of the rapidity Y, for $B/r_0=10$. }}
\label{fig:on1}
\end{figure} 

\begin{figure}
\begin{center}
  \includegraphics[angle=270, scale=0.7]{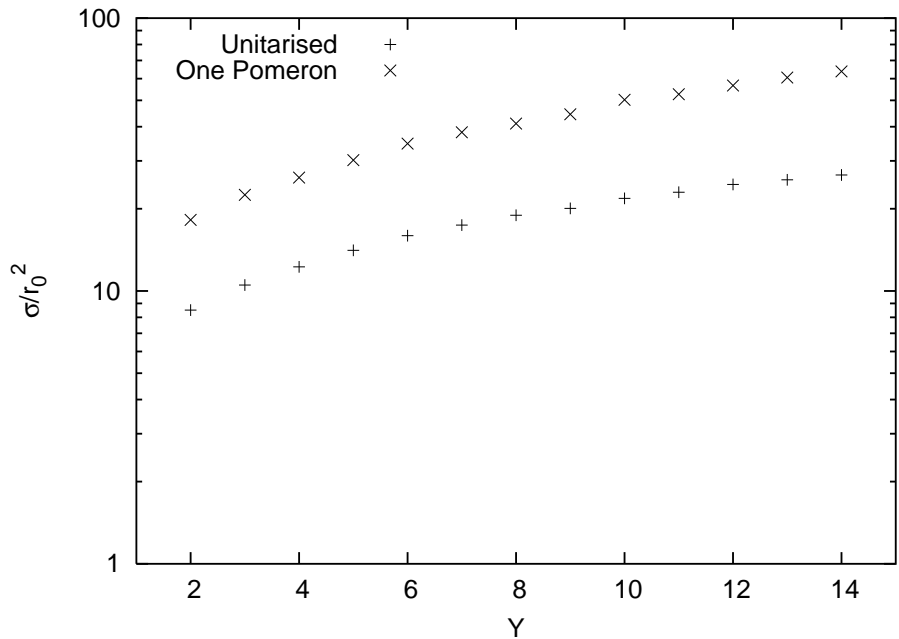}
\end{center}
\caption{\emph{$\sigma/r_0^2$ for unitarised and one pomeron calculations for onium-nucleus scattering as
function of the rapidity Y, for $B/r_0=5$. }}
\label{fig:on2}
\end{figure}

Figures \ref{fig:on1} and \ref{fig:on2} show the results for onium-nucleus scattering. The saturation effects are more visible here. For small $Y$ we see the power like rise of $\sigma$ and as $Y$ increases the cross section starts to saturate. The reason that we can see the difference between the one pomeron and the unitarised amplitudes even for small $Y$ is because the onium now interacts with a much denser object. Observe also that the $Y$ in onium-onium collisions and the $Y$ in onium-nucleus scattering has different meaning for the evolution of an onium state. In onium-onium scattering events, the onium states are evolved up to $Y/2$ while in onium-nucleus scattering, the state is evolved up to $Y$. For a low rapidity, such as $Y=2$ ($Y=4$ in onium-onium events) the onium state often consists only of the initial dipole, therefore an onium-onium collision will just mean a collision between two dipoles. Hence it will be very difficult to see any difference between the one pomeron and unitarised amplitudes. In nucleus collisions however, that single dipole will collide with a much denser object, which consists of many dipoles, and therefore the amplitudes will not be so small and the difference will be more visible. There is a difference here compared to the case of constant $\rho$. If one uses a constant $\rho$ , which is sufficiently small of course, the steps in $y$ will be smaller and therefore there will always be additional dipoles even for smaller $Y$. Hence the difference in the amplitudes will be visible, though small. This can be seen in \cite{e9}.  

\section{More on Saturation}

\subsection{The Dipole Fusion Factor}

In this section we will focus on a process that we excluded in our analysis, namely the question 
of dipole fusion processes, the creation of a dipole by destroying two. To make the analysis easier we abandon energy conservation and switch to a constant cut-off, $\rho$. We mentioned the fusion process a 
little when we presented the Balitsky-Kovchegov equation,~\eqref{eq:BK}. As we said, the BK equation 
does not contain this possibility. This is so because Mueller's Dipole 
formulation does not contain it, and the BK equation is obtained from~\eqref{eq:mueller} 
by summing multiple pomeron exchanges up to all orders. The inclusion of dipole fusion processes in the wave function formalism is not an easy task and is yet to be solved. 
 
Let us return to Mueller's formulation. As we recall from section 2, Mueller's model  
gives us the squared wave function for an onium state with arbitrarily many gluons. We  
also saw that, when two new dipoles, $in$ and $jn$, were created as the result of a gluon  
emission from dipole $ij$, there was a factor $\bar{\bar{\alpha}}d^2x_n\frac{x_{ij}^2}{x_{in}^2x_{jn}^2}$ associated 
with this splitting. The wave function, $\phi^{(n)}$, is then obtained from $\phi^{(0)}$ by considering all possible ways in which we can arrive at a state with $n$ gluons. 
\begin{figure}
\begin{center}
\begin{picture}(400,100)(0,0)
\LongArrow(0,50)(50,50)
\Vertex(110,50){2}
\Vertex(310,50){2}
\Line(0,100)(70,75)
\Line(0,0)(70,25)
\Line(70,75)(70,25)
\Line(70,25)(110,50)
\Line(70,25)(110,50)
\Line(110,50)(150,75)
\Line(110,50)(150,25)
\Line(70,75)(150,75)
\Line(70,25)(150,25)
\Line(70,75)(110,50)
\Text(110,5)[]{Before}
\LongArrow(200,50)(250,50)
\Line(200,100)(270,75)
\Line(200,0)(270,25)
\Line(270,75)(270,25)
\Line(270,25)(310,50)
\Line(270,25)(310,50)
\Line(310,50)(350,75)
\Line(310,50)(350,25)
\Line(270,75)(350,75)
\Line(270,25)(350,25)
\Line(270,75)(310,50)
\Text(310,5)[]{After}
\Line(350,75)(350,25)
\LongArrow(362,50)(400,50)
\Text(95,64)[]{$a$}
\Text(95,36)[]{$b$}
\Text(135,62)[]{$c$}
\Text(135,40)[]{$d$}
\Text(110,20)[]{$e$}
\Text(110,80)[]{$f$}
\Text(65,50)[]{$r_0$}
\Text(295,64)[]{$a$}
\Text(295,36)[]{$b$}
\Text(335,62)[]{$c$}
\Text(335,40)[]{$d$}
\Text(310,20)[]{$e$}
\Text(310,80)[]{$f$}
\Text(265,50)[]{$r_0$}
\Text(357,50)[]{$r_1$}
\end{picture}
\end{center}
\caption{\emph{A dipole fusion process. The dipoles denoted by c and d go together and form $r_1$ and the evolution continues}}
\label{fusion}
\end{figure}
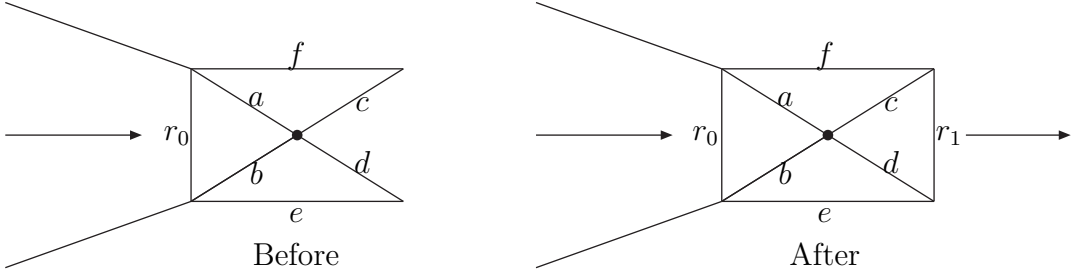

Look at figure \ref{fusion}, where we follow some evolution that leads to the formation of the dipole labeled by $r_0$. Then $r_0$ splits into $a$ and $b$  , where $a$ in turn splits into $f$ and $c$, and $b$ splits into $d$ and $e$. Then, from what we said above, we will get a factor (denote the gluon at the center with $A$)
\begin{equation}
d^2r_A\frac{r_0^2}{a^2b^2}\frac{a^2}{f^2c^2}\frac{b^2}{d^2e^2}
\label{eq:fusion2}
\end{equation}
for these splittings (of course we will also get factors, $d^2r$, from the gluons on the top right and bottom right corners, but these factors are not interesting for our analysis). Note that in the large $N_c$ limit only dipoles that share a common gluon, i.e neighboring dipoles, can fuse. After these splittings we let $c$ and $d$ fuse and form $r_1$. The fusion process will be associated with some weight $\mathcal{Z}\cdot dy$, and the question is what this factor looks like. Observe also that we do not need to write $r_0^2$ in~\eqref{eq:fusion2} since there will be a factor $1/r_0^2$ from the time $r_0$ was formed and therefore these factors cancel. Instead we can include a factor $r_1^2$ which will be present when the dipole $r_1$ emits a gluon, later in the process. Thus we get
\begin{equation}
d^2r_A\frac{1}{f^2c^2}\frac{1}{d^2e^2}\mathcal{Z}(r_1;c,d)r_1^2
\label{eq:fusion12}
\end{equation}
where the arguments of $\mathcal{Z}$ indicate that the dipole $r_1$ is formed out of the dipoles $c$ and $d$. As discussed above, we require that the theory looks the same if we view the process from the other end. If we start from the right and follow the evolution we will see $r_1$ form first, then $r_1$ will split into $d$ and $c$, $c$ will in turn split into $a$ and $f$ while $d$ splits into $b$ and $e$, and finally $a$ and $b$ go together to create $r_0$ and the process continues. Now we will get the factor
\begin{equation}
d^2r_A\frac{r_1^2}{c^2d^2}\frac{c^2}{f^2a^2}\frac{d^2}{b^2e^2}\mathcal{Z}(r_0;a,b)
\end{equation}

Applying the same arguments as above we get rid of the factor $r_1^2$ and we can include $r_0^2$. Hence we obtain
\begin{equation}
d^2r_A\frac{1}{f^2a^2}\frac{1}{b^2e^2}\mathcal{Z}(r_0;a,b)r_0^2
\label{eq:fusion22}
\end{equation}
By the symmetry argument~\eqref{eq:fusion12} and~\eqref{eq:fusion22} should be equal, hence
\begin{eqnarray}
\frac{r_0^2}{a^2b^2e^2f^2}\mathcal{Z}(r_0;a,b)&=&\frac{r_1^2}{c^2d^2e^2f^2}\mathcal{Z}(r_1;c,d) \nonumber \\
\frac{r_0^2}{a^2b^2}\mathcal{Z}(r_0;a,b)&=&\frac{r_1^2}{c^2d^2}\mathcal{Z}(r_1;c,d)
\label{eq:fusionn3}
\end{eqnarray}

Most cascade models are semi-classical but despite this they are very successful in describing experimental results, such as in $e^+e^-$ annihilation. The DGLAP region in DIS is also successfully described despite the fact that DGLAP, which contains probabilities instead of amplitudes, is not fully quantum mechanical. In this, semi-classical, spirit we assume that the fusion factor, just like the splitting factor, is local and only depends on the dipoles involved in the fusion. A full quantum mechanical treatment would give contributions which include interference factors but, given the success of these semi-classical models, we will primarily try a locally factorizing approximation. The assumption that the fusion factor is local means that $\mathcal{Z}(r_0;a,b)$ ($\mathcal{Z}(r_1;c,d)$) only contains the lengths $r_0, a$ and $b$ ($r_1, c$ and $d$). Thus the only possibility gives 
\begin{equation}
\mathcal{Z}(r_0;a,b)\propto \frac{a^2b^2}{r_0^2} \; \; \textrm{and}\; \; \mathcal{Z}(r_1;c,d)\propto \frac{c^2d^2}{r_1^2}
\label{eq:fusion3}
\end{equation}     
Generally, if dipoles $r_{in}$ and $r_{jn}$ go together and form $r_{ij}$ we get a factor $\frac{r_{in}^2r_{jn}^2}{r_{ij}^2}$. To make the formulas in~\eqref{eq:fusion3} complete we note that the fusion factor $\mathcal{Z}$ must be dimensionless. Therefore the expressions in~\eqref{eq:fusion3} should be divided by a factor which has the dimension of an area. A reasonable choice would be to take the area of the triangle that is formed by the two disappearing dipoles and the dipole that is created. If this is done however, we get a factor $1/\mathcal{A}_1$ on the LHS of~\eqref{eq:fusionn3} while we get $1/\mathcal{A}_2$ on the RHS, where $\mathcal{A}_1$ is the area of triangle formed by $a$, $b$ and $r_0$ while $\mathcal{A}_2$ is the area of the triangle $r_1cd$.
Since in general $\mathcal{A}_1\neq\mathcal{A}_2$, the equality, hence the symmetry, will be violated. In order to keep the symmetry one can easily see that we must have something that looks like $1/\sqrt{\mathcal{A}_1\mathcal{A}_2}$. Including $\mathcal{A}_1$ ($\mathcal{A}_2$) in the fusion factor, $\mathcal{Z}(r_1;c,d)$ ($\mathcal{Z}(r_0;a,b)$),  means that we remember the steps that we took before. From what we said above however, we avoid this possibility and assume that the fusion factor is local. Therefore $\mathcal{Z}(r_1;c,d)$ ($\mathcal{Z}(r_0;a,b)$) will not contain $\mathcal{A}_1$ ($\mathcal{A}_2$), but as we just saw we cannot get a theory which is symmetric by only including $\mathcal{A}_2$ ($\mathcal{A}_1$). Thus the only possibility is that the fusion factor contains a parameter, which has the dimension $(length)^2$, and is a fundamental parameter in the theory. One can ask what kind of fundamental scales there are. The fundamental scale of QCD is $\Lambda_{QCD}$ which has the dimension of energy. Therefore one can make the heuristic argument that $\mathcal{Z}\propto \Lambda_{QCD}^2$. It is also reasonable to assume that the fusion process, just like the splitting process, is associated with a factor $\bar{\bar{\alpha}}$. The complete formula for the fusion of dipoles $in$ and $jn$ into $ij$ is then given by
\begin{equation}
\mathcal{Z}(r_{ij};r_{in},r_{jn})=\bar{\bar{\alpha}}\xi \frac{r_{in}^2r_{jn}^2}{r_{ij}^2}\Lambda_{QCD}^2
\label{eq:fusion4}
\end{equation}
where $\xi$ is a dimensionless, free parameter in the theory.  
    
\subsection{Putting the Fusion Factor in a MC Program}

Now that we have the fusion factor we can use it to generate new $y$ values, which we denote by $\hat{y}$. As mentioned in section 3 we generate, for each existing dipole, a $y$ value and then we accept the gluon with the lowest $y$-value. For each dipole we also generate $\hat{y}$-values for possible dipole fusions and if the lowest $\hat{y}$ happens to be lower than all the other $y$ we accept the corresponding fusion process. The $\hat{y}$-values are given by
\begin{equation}
\hat{y}=y-\frac{1}{\bar{\bar{\alpha}}\xi}\frac{1}{\Lambda_{QCD}^2}\frac{r_{ij}^2}{ r_{in}^2r_{jn}^2}\textrm{ln}R,
\label{eq:raphat}
\end{equation}
where $R$ is a random number and $y$ is the maximum rapidity before the fusion process. 
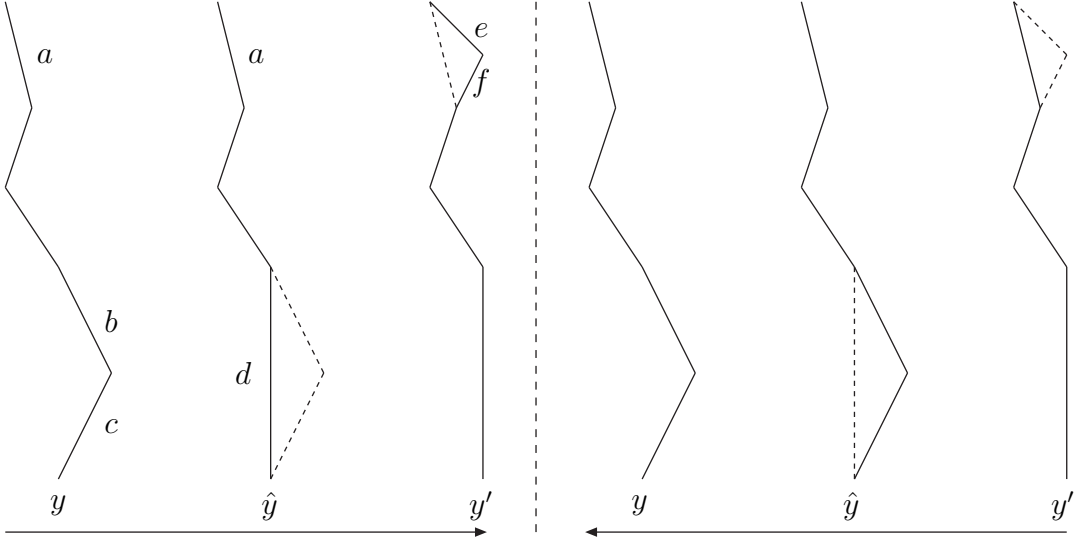
\begin{figure}
\begin{center}
\begin{picture}(400,200)(0,0)
\Line(0,200)(10,160)
\Line(10,160)(0,130)
\Line(0,130)(20,100)
\Line(20,100)(40,60)
\Line(40,60)(20,20)
\Line(80,200)(90,160)
\Line(90,160)(80,130)
\Line(80,130)(100,100)
\Line(100,100)(100,20)
\DashLine(100,100)(120,60){2}
\DashLine(120,60)(100,20){2}
\DashLine(160,200)(170,160){2}
\Line(160,200)(180,180)
\Line(180,180)(170,160)
\Line(170,160)(160,130)
\Line(160,130)(180,100)
\Line(180,100)(180,20)
\Text(20,10)[]{$y$}
\Text(100,10)[]{$\hat{y}$}
\Text(180,10)[]{$y'$}
\LongArrow(0,0)(180,0)
\Line(220,200)(230,160)
\Line(230,160)(220,130)
\Line(220,130)(240,100)
\Line(240,100)(260,60)
\Line(260,60)(240,20)
\Line(300,200)(310,160)
\Line(310,160)(300,130)
\Line(300,130)(320,100)
\Line(320,100)(340,60)
\Line(340,60)(320,20)
\DashLine(320,100)(320,20){2}
\DashLine(380,200)(400,180){2}
\DashLine(400,180)(390,160){2}
\Line(380,200)(390,160)
\Line(390,160)(380,130)
\Line(380,130)(400,100)
\Line(400,100)(400,20)
\Text(240,10)[]{$y$}
\Text(320,10)[]{$\hat{y}$}
\Text(400,10)[]{$y'$}
\LongArrow(400,0)(220,0)
\DashLine(200,0)(200,200){3}
\Text(15,180)[]{$a$}
\Text(40,80)[]{$b$}
\Text(40,40)[]{$c$}
\Text(90,60)[]{$d$}
\Text(95,180)[]{$a$}
\Text(180,190)[]{$e$}
\Text(180,170)[]{$f$}
\end{picture}
\end{center}
\caption{\emph{On the left side we see $b$ and $c$ fuse to create $d$ at a rapidity $\hat{y}$ and then $a$ splits into $e$ and $f$ at $y'$. The figure on the right side shows the same process viewed backward. Now $e$ and $f$ undergo fusion to create $a$ at $y'$ and $d$ splits into $b$ and $c$ at a rapidity $\hat{y}$.}}
\label{yhat}
\end{figure}

One may ask what physical meaning $\hat{y}$ has. Consider the processes on figure \ref{yhat}, where we picture the evolution of an onium state for some rapidity intervals and then we view the same steps backward. When we come to $y$ we generate and choose $\hat{y}$ and let $b$ and $c$ go together to form $d$. When viewed backward the meaning of $\hat{y}$ will be the rapidity at which $d$ splits into $b$ and $c$. The same thing goes for the process where $a$ splits into $e$ and $f$ at $y'$. Viewing the process backward we see $e$ and $f$ undergo fusion and create $a$ and the rapidity $y'$ would then be the rapidity which we would have obtained from~\eqref{eq:raphat} for the backward evolution. 

\subsection{Stability}

Let us return to figure \ref{fusion} and the gluon at the center denoted by A.  What happens if we vary $\bold{r}_A$? In particular consider events where $A$ is emitted far away from the mother dipole. After the fusion of $c$ and $d$ into $r_1$ there will be no terms left that depend on $\bold{r}_A$. Therefore, if we integrate $d^2r_A$ over large distances we will get an infinite contribution. The problem is that even though the probability to emit the gluon $A$ at very large distances goes to zero, the probability of having the fusion event $c+d\rightarrow r_1$ goes to infinity. This can be easily seen from the fusion factor $\frac{c^2d^2}{r_1^2}$. Hence in order to make the theory stable the contribution from very large dipoles must be suppressed. Enforcing energy conservation we got a cut-off for the large dipoles from $p_-$ conservation but in the case of constant $\rho$ there is no limit for how large a dipole can be. It should be reasonable to use an infrared cut-off even if one does not consider fusion processes. This has to do with the Froissart bound, which says that the cross section cannot grow faster than ln$^2s$. The Froissart bound follows from the assumption that a field, which acts with the exchange of massive particles, has a finite range. Therefore one should cut off contributions that come from too large distances, which means that it is not right to allow too large dipoles. In QCD this is expressed in the confinement mechanism, which suppresses large dipoles. A reasonable infrared cutoff is given by $1/\Lambda_{QCD}$. In a Monte Carlo program the easiest way to implement this would be to generate the $y$, $r_x$ and $r_y$ as before and afterwards accept these values with a probability, $\mathcal{P}\sim e^{-(r_1+r_2)\Lambda_{QCD}}$. Here $r_1$ and $r_2$ are the sizes of the dipoles which were created in the splitting process.     

Next, let us study what happens when the gluon $A$ is emitted very close to one of its mother partons. Denote the smaller distance to the mother partons by $\delta$. Then for very small $\delta$, the probability to emit $A$ gets very large but in order to have a fusion process, involving a dipole that contains $A$, we must first have an emission from the dipole with length $\delta$, which then will get a weight proportional to $\delta^2$. Therefore the total probability for the whole process will not be large, and since the value of $\delta$ varies over a small space we will not get a divergent contribution. Therefore we see that including fusion processes causes no unpleasant instabilities in the theory. 

One can ask how large effect the fusion processes will have on the onium evolution. Since the probability to create small dipoles is high (of course we assume that $\rho$ is small) there will be a lot of small dipoles. On the other hand, the probability to have fusion processes involving small dipoles is very low and therefore it would seem as if the fusion effects would not be so visible. Small dipoles interact very weakly however, therefore it does not make so much difference if many of them disappear due to fusion processes or not. The largest contribution to interactions come from the larger dipoles and these have non-vanishing probabilities to undergo fusion, therefore for a state with many large dipoles the effects of fusion processes should be visible.

\subsection{Results}

In this section we present the results we obtained from our MC simulations including dipole fusion processes. We have run the program for different $\xi$ and studied how the cross section behaves for these $\xi$-values, in onium-onium collisions.   
\begin{figure}
\begin{center}
  \includegraphics[angle=270, scale=0.7]{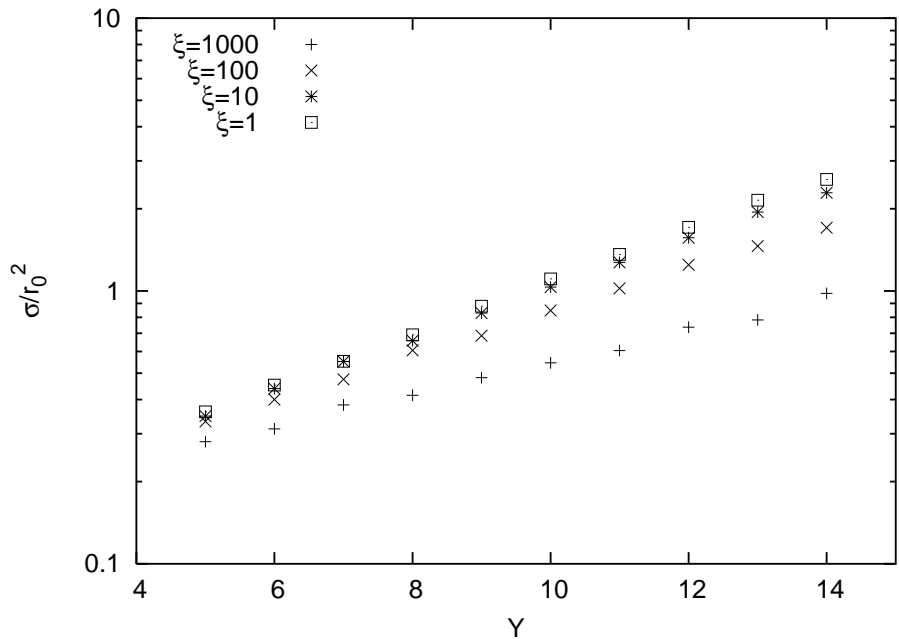}
\end{center}
\caption{$\sigma/r_0^2$ plotted as a function of $Y$ for various $\xi$ and for constant $\rho=0.02r_0$.}
\label{puref}
\end{figure}
In figure \ref{puref} we see some of the result obtained by including fusion effects. We choose $r_0$ such that $r_0\Lambda_{QCD}=0.166$. The cross sections were calculated for fix $\rho$. We see that $\lambda$ decreases as $\xi$ increases and ln$\sigma$ still seems to depend linearly on $Y$. It doesn't seem as if the growth starts to saturate but of course it might be that the saturation sets in for larger $Y$. This is actually expected in the large $N_c$ limit where only neighboring dipoles can fuse. From these result we have also calculated the different $\lambda$ and in figure \ref{lambdaxi}, $\lambda$ is plotted as a function of $\xi$. 

Let $\lambda_F$ denote the $\lambda$-value we obtain from onium evolution including fusion processes.
\begin{figure}
\begin{center}
  \includegraphics[angle=270, scale=0.7]{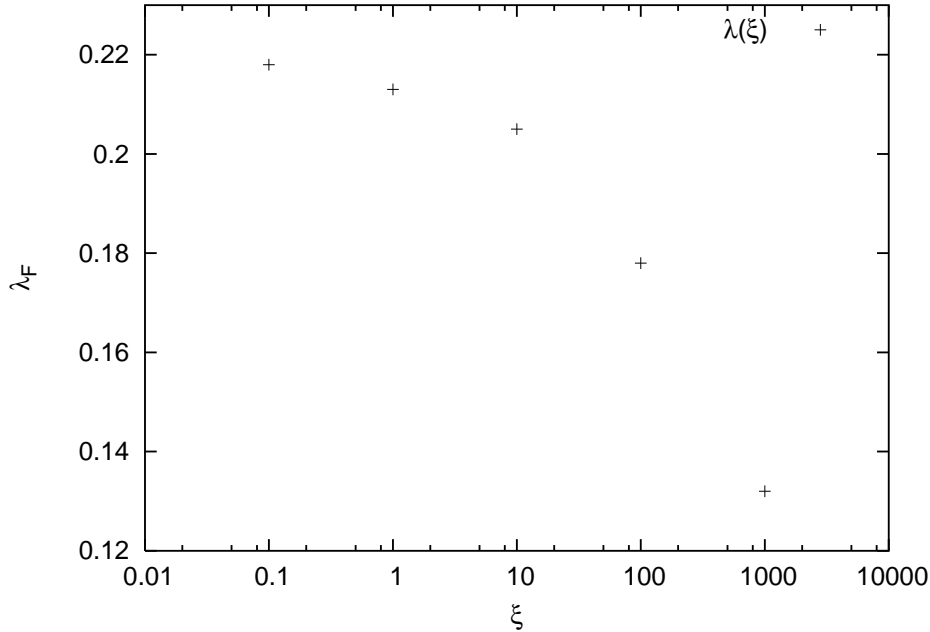}
\end{center}
\caption{$\lambda$ plotted as a function of $\xi$.}
\label{lambdaxi}
\end{figure}
Similarly we introduce $\lambda_{EC}$, where EC stands for energy conservation. From figure \ref{lambdaxi} we see that $\lambda_F$ drops slowly for small $\xi$. This is not strange since for small $\xi$ ($\xi < 1$), any fusion process will have very a small probability to occur, and changing $\xi$ will not affect this much. As $\xi$ increases we see that $\lambda_F$ starts drop more sharply since the fusion processes become more probable. We have not gone any further than $\xi=1000$ since running the program for larger $\xi$ is extremely time consuming. The reason is that the steps in $y$ gets extremely small when $\xi$ is very large, which follows immediately from~\eqref{eq:raphat}. 

Looking at the figures \ref{fig:lambda} and \ref{lambdaxi} we see that $\lambda_F >\lambda_{EC}$. This means that energy conservation has more effect on the growth of $\sigma$ than fusion processes. It can be that the fusion processes become more important than energy conservation at higher rapidities. Of course when estimating $\lambda_{EC}$ we used values within the interval $10\leqslant Y\leqslant 19$ but for $\lambda_F$ we used values between $Y=8$ and $Y=14$. The results would be slightly modified if one uses this interval to determine $\lambda_{EC}$. For the lower $Y$ range we would, for $r_1=r_2$, find $\lambda_{EC}=0.12$ instead of $\lambda_{EC}=0.10$. The new $\lambda_{EC}$ still satisfies $\lambda_{EC} < \lambda_F$ though. 

If we calculate the cross section without fusion processes but imposing the infrared cut off we get $\lambda\approx0.21$. This might seem strange since $\lambda_F$ for $\xi=0.1$ is approximately equal to $0.22$. This is not surprising though since there are some uncertainties in the $\lambda$-values. Also, $\lambda_F(\xi=0.1)$ should more or less be equal to $\lambda$, since any fusion process will be highly unlikely for these values. Especially when we have an infrared cut off. Finally we note that the fusion processes have little impact on $\lambda$ for $\xi \lesssim 10$.       

\section{Outlook}     

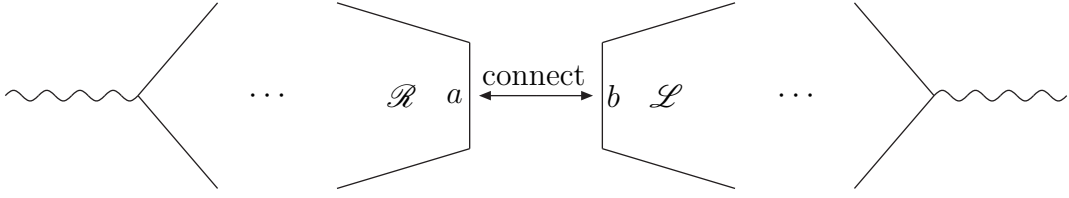
\begin{figure}
\begin{center}
\begin{picture}(400,70)(0,0)
\Photon(0,35)(50,35){2}{4}
\Photon(400,35)(350,35){2}{4}
\Line(50,35)(80,70)
\Line(50,35)(80,0)
\Line(350,35)(320,70)
\Line(350,35)(320,0)
\Line(125,70)(175,55)
\Line(125,0)(175,15)
\Line(175,55)(175,15)
\Line(275,70)(225,55)
\Line(275,0)(225,15)
\Line(225,55)(225,15)
\LongArrow(200,35)(180,35)
\LongArrow(200,35)(220,35)
\Text(170,35)[]{$a$}
\Text(230,35)[]{$b$}
\Text(150,35)[]{$\mathscr{R}$}
\Text(250,35)[]{$\mathscr{L}$}
\Text(200,43)[]{connect}
\Text(100,35)[]{$\cdots$}
\Text(300,35)[]{$\cdots$}
\end{picture}
\end{center}
\caption{\emph{The two states are connected between dipoles $a$ and $b$. The dots just indicate that there has been an evolution up to $a$ and $b$ and $\mathscr{R}$ ($\mathscr{L}$) denote the right (left) moving onium.} }
\label{connect}
\end{figure}

\begin{figure}
\begin{center}
\begin{picture}(400,70)(0,0)
\Photon(0,35)(50,35){2}{4}
\Photon(400,35)(350,35){2}{4}
\Line(350,35)(320,70)
\Line(350,35)(320,0)
\Line(50,35)(80,70)
\Line(50,35)(80,0)
\Line(125,70)(175,55)
\Line(125,0)(175,15)
\Line(175,55)(175,15)
\Line(275,70)(225,55)
\Line(275,0)(225,15)
\Line(225,55)(225,15)
\Line(175,55)(225,55)
\Line(175,15)(225,15)
\Line(175,55)(225,15)
\Line(175,15)(225,55)
\Text(170,35)[]{$a$}
\Text(230,35)[]{$b$}
\Text(200,60)[]{$e$}
\Text(200,7)[]{$f$}
\Text(185,30)[]{$c$}
\Text(215,30)[]{$d$}
\Text(150,35)[]{$\mathscr{R}$}
\Text(250,35)[]{$\mathscr{L}$}
\Text(100,35)[]{$\cdots$}
\Text(300,35)[]{$\cdots$}
\end{picture}
\end{center}
\caption{\emph{In this figure, $c$ and $d$ are the diagonal distances between $a$ and $b$.}}
\label{cd}
\end{figure}
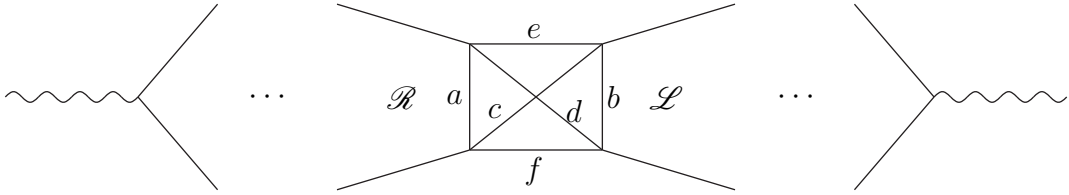

In this section we will present three problems that can be investigated in future studies. The first problem is certainly something very interesting. Consider two onia states that we connect at some rapidity, see figure \ref{connect}. Denote the right (left) moving onium with $\mathscr{R}$ ($\mathscr{L}$). Assume that the interaction between the two states occurs between the dipoles $a$ and $b$, where $a$ ($b$) belong to $\mathscr{R}$ ($\mathscr{L}$).  
The formula we used for the interaction between two dipoles is given by~\eqref{eq:salam}. This formula can be written in the form 
\begin{equation}
\frac{\alpha_s^2}{2}\textrm{ln}^2\frac{cd}{ef}=\frac{\alpha_s^2}{8}\textrm{ln}^2\frac{c^2d^2}{e^2f^2}
\label{eq:emil}
\end{equation}
For the meaning of $c$, $d$, $e$ and $f$, see figure \ref{cd}. Observe that figures 22 and 23 are extremely simplified, just to give an idea of what happens. In general it is highly unlikely that two onia collide head on and the dipoles $a$ and $b$ can have any relative orientation.

The expression above does not factorize in the same way as the rest of the splitting factors in the chain. However, if the  
 separation between the dipoles is large compared to their sizes it does reproduce an expression that is similar to the splittings factors. Thus $e$ and $f$ are much larger than $a$ and $b$. We can choose the vectors such that $\bold{c}=\bold{a}+\bold{e}$, $\bold{d}=\bold{e}+\bold{b}$ and $\bold{f}=\bold{a}+\bold{e}+\bold{b}$. Then we have
\begin{eqnarray}
c^2&=&a^2+e^2+2ae\cos \theta \nonumber \\
d^2&=&b^2+e^2+2be\cos \phi \nonumber \\
f^2&=&a^2+b^2+e^2+2ae\cos \theta + 2be\cos \phi +2ab\cos \psi
\end{eqnarray}
where, using our definitions of the vectors, we have $\psi = \theta - \phi$. Of course one can choose the orientations such that  $\psi = \theta + \phi$ also, but this will not change the result. Next, we define $\epsilon = \frac{a}{e}$ and $\delta = \frac{b}{e}$. Then we get
\begin{eqnarray}
\frac{c^2}{e^2}&=&1+\epsilon^2+2\epsilon \cos \theta \nonumber \\
\frac{d^2}{e^2}&=&1+\delta^2+2\delta \cos \phi \nonumber \\
\frac{f^2}{e^2}&=&1+\epsilon^2+\delta^2+2\epsilon \cos \theta + 2\delta \cos \phi +2\epsilon \delta \cos \psi
\end{eqnarray}
Using these relations we obtain 
\begin{eqnarray}
\textrm{ln}\frac{c^2}{e^2}&\approx& \epsilon^2+2\epsilon \cos \theta -\frac{1}{2}(\epsilon^2+2\epsilon \cos \theta)^2\nonumber \\
\textrm{ln}\frac{d^2}{e^2}&\approx& \delta^2+2\delta \cos \phi -\frac{1}{2}(\delta^2+2\delta \cos \phi)^2\nonumber \\
\textrm{ln}\frac{f^2}{e^2}&\approx& \epsilon^2+\delta^2+2\epsilon \cos \theta + 2\delta \cos \phi + 2\epsilon \delta \cos \psi - \nonumber \\
& &-\frac{1}{2}(\epsilon^2+\delta^2+2\epsilon \cos \theta + 2\delta \cos \phi + 2\epsilon \delta \cos \psi)^2
\end{eqnarray}
Therefore we get
\begin{eqnarray}
\textrm{ln}\frac{c^2d^2}{e^2f^2}&=&\textrm{ln}\frac{c^2}{e^2}+\textrm{ln}\frac{d^2}{e^2}-\textrm{ln}\frac{f^2}{e^2} \nonumber \\
&\approx&-2\epsilon \delta \cos \psi + 2\epsilon^2 \delta^2 \cos \psi + \epsilon^2 \delta^2+2\epsilon^2 \delta \cos \phi+ 2\epsilon^3 \delta \cos \psi + 2\epsilon \delta^3 \cos \psi + \nonumber \\
&+& 2\epsilon \delta^2 \cos \theta + 4\epsilon \delta \cos \phi \cos \theta+ 4\epsilon^2 \delta \cos \theta \cos \psi + 4\epsilon \delta^2 \cos \psi \cos \phi \nonumber \\
&\approx& 4\epsilon \delta \cos \phi \cos \theta -2\epsilon \delta \cos \psi 
\end{eqnarray}
 where we have kept terms to second  order in $\epsilon$ and $\delta$. Using the relation $\psi = \theta - \phi$ we get
\begin{equation}
\textrm{ln}\frac{c^2d^2}{e^2f^2}\approx 2\epsilon \delta (\cos \phi \cos \theta-\sin \phi \sin \theta)
\end{equation} 
Squaring this expression we get
\begin{equation}
4\epsilon^2 \delta^2 (\cos^2 \phi \cos^2 \theta-2\cos \phi \sin \theta \cos \phi \sin \theta + \sin^2 \phi \sin^2 \theta)
\end{equation}
If we consider all the possible orientations of $\bold{a}$ and $\bold{b}$, while keeping $\bold{e}$ fixed, we should take the average of the expression above
\begin{equation}
4\epsilon^2 \delta^2 \langle \cos^2 \phi \cos^2 \theta-2\cos \phi \sin \theta \cos \phi \sin \theta + \sin^2 \phi \sin^2 \theta \rangle_{\theta,\phi} 
= 4\epsilon^2 \delta^2(\frac{1}{4}+\frac{1}{4})
\end{equation}
Hence we have
\begin{equation}
\frac{\alpha_s^2}{8}\textrm{ln}\frac{c^2d^2}{e^2f^2}\approx \frac{\alpha_s^2}{8}2\epsilon^2\delta^2\approx \biggl(\frac{\alpha_s}{2}\biggr)^2\frac{a^2b^2}{e^2f^2} \sim \bar{\bar{\alpha}}^2\frac{a^2b^2}{e^2f^2}
\end{equation}
Thus with the approximation that the separation bewteen the interacting dipoles is much larger than their sizes we can write the dipole-dipole amplitude as
\begin{equation}
 \bar{\bar{\alpha}}^2\frac{a^2b^2}{e^2f^2}
\label{eq:amplitude}
\end{equation}

Let us return to the onia states in figures \ref{connect} and \ref{cd}. It is easy to see that, using~\eqref{eq:emil}, we do not get the same result if we connect the states at different rapidities. This is somewhat disturbing, we would like to have an amplitude such that we can connect the onia states at any given rapidity and get the same results. This would be true if the total weight for the evolution of the states $\mathscr{R}$ and $\mathscr{L}$, multiplied with the amplitude between $a$ and $b$ would be equal to the weight of a single chain, which is spanned between the ends. The weight for a chain is simply the product of all the splitting factors (to make things simpler we assume no fusion processes). To make the weight symmetric the reader can easily verify that we must have an amplitude that is equal to~\eqref{eq:amplitude}. 
 
It would be interesting to find out what the splitting factors would be, if one wants to have the same symmetry by using the amplitude~\eqref{eq:emil}. Obviously such a splitting factor would be similar to~\eqref{eq:emil}. It would perhaps follow from next to leading order calculations, which is a better approximation of Mueller's model. Then, just as we did in deriving~\eqref{eq:amplitude}, it would be possible that one obtains the splitting factor in Mueller's model as an approximation of the new splitting factor. In this new splitting factor, the contributions from the different dipoles would not factorize as before, but it would contain different interference terms, which can be seen from~\eqref{eq:emil}. 

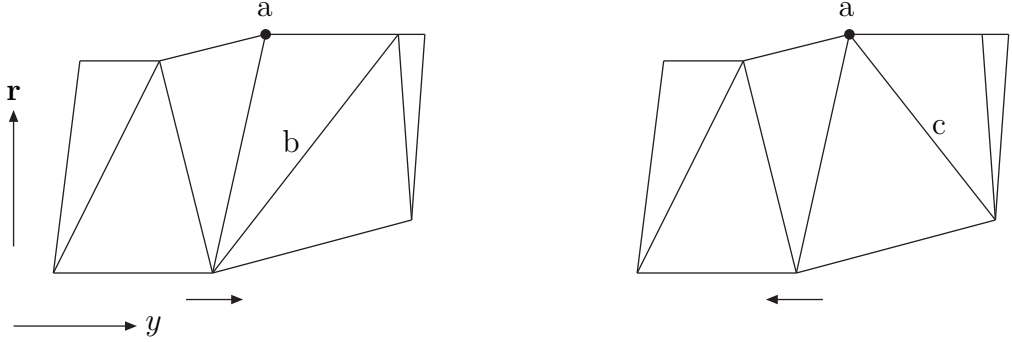
\begin{figure}
\begin{center}
\begin{picture}(400,150)(0,0)
\Line(20,20)(30,100)
\Line(30,100)(60,100)
\Line(20,20)(60,100)
\Line(80,20)(20,20)
\Line(60,100)(80,20)
\Line(100,110)(80,20)
\Line(60,100)(100,110)
\Line(100,110)(150,110)
\Line(150,110)(80,20)
\Line(150,110)(155,40)
\Line(80,20)(155,40)
\Line(155,40)(160,110)
\Line(150,110)(160,110)
\Vertex(100,110){2}
\Text(100,120)[]{a}
\LongArrow(70,10)(90,10)
\Line(240,20)(250,100)
\Line(250,100)(280,100)
\Line(240,20)(280,100)
\Line(300,20)(240,20)
\Line(280,100)(300,20)
\Line(320,110)(300,20)
\Line(280,100)(320,110)
\Line(320,110)(370,110)
\Line(320,110)(375,40)
\Line(370,110)(375,40)
\Line(300,20)(375,40)
\Line(375,40)(380,110)
\Line(370,110)(380,110)
\Vertex(320,110){2}
\Text(320,120)[]{a}
\LongArrow(310,10)(290,10)
\Text(110,70)[]{b}
\Text(355,75)[]{c}
\LongArrow(5,30)(5,80)
\Text(5,88)[]{$\bold{r}$}
\LongArrow(5,0)(50,0)
\Text(58,0)[]{$y$}
\end{picture}
\end{center}
\caption{\emph{On the left figure we start from the left and evolve towards the right while on the right figure we the evolution starts at the right end and stops on the left}}
\label{RL}
\end{figure}

We leave the first problem and go on to the second. As mentioned earlier we have tried to construct a theory that is right-left symmetric. The weights we use are symmetric but everything is not exactly symmetric. Consider figure \ref{RL}, where the rapidity increases to the right. Recall from section 3.2 that the transverse momentum of a gluon is decided by the shortest distance to another gluon, with which it has formed a dipole.  

We look closer at the gluon denoted by $a$, as we can see this gluon has different links connecting it to the other gluons in the two figures. Therefore it can have different $p_\perp$ values depending on which side we start from. When we evolve from left to right the dipole denoted by $b$ forms at some stage but when we start from right, the same dipole never forms, instead we get the dipole $c$ which is not seen on the left figure. Observe that the weights will be the same in both the right and the left figures since there will be no factors depending on $b$ and $c$. Therefore it is not a big problem that we do not have the $p_\perp$ symmetry. An improvement would be to make a better approximation, in choosing the $p_\perp$ values for an emission, that respects $\bold{p}_\perp$ conservation. Another, and a very natural, improvement would of course be to use a running coupling constant $\alpha$, which can be an interesting project for the future. 

Finally, we look at the third problem  which might also be interesting.  Going back to the BK equation,~\eqref{eq:BK}, we mentioned that this equation was derived from~\eqref{eq:mueller} which uses a constant ultraviolet cut-off $\rho$. It would be interesting to try to modify the BK equation with a energy conserving, i.e $y$ dependent, cut-off. One obvious change would be in the Sudakov factors of~\eqref{eq:mueller}. As we mentioned in section 2.2 the first Sudakov factor of~\eqref{eq:mueller} described the probability that nothing happens during the evolution, from $y=0$ to $y=Y$. This Sudakov factor has the form
\begin{equation}
S=exp\biggl[-\bar{\bar{\alpha}}\int_0^Y dy\int_\rho d^2r_2\frac{r_{10}^2}{r_{12}^2r_{20}^2}\biggr]=exp\biggl[-2\bar{\alpha}\textrm{ln}\biggl(\frac{r_{10}}{\rho}\biggr)Y\biggr]
\end{equation}
for constant $\rho$. If we make the cut-off y dependent we get
\begin{eqnarray}
S&=&exp\biggl[-\bar{\bar{\alpha}}\int_0^Y dy\int_{\rho (y)} d^2r_2\frac{r_{10}^2}{r_{12}^2r_{20}^2}\biggr]=exp\biggl[-2\bar{\alpha}\int_0^Y dy\textrm{ln}\biggl(\frac{r_{10}}{\rho (y)}\biggr)\biggr] \nonumber \\
&=&exp\biggl[-2\bar{\alpha}\int_0^Y dy(\textrm{ln}(r_{10}p_+) + y)\biggr]=exp\biggl[-2\bar{\alpha}(\textrm{ln}(r_{10}p_+)Y + \frac{Y^2}{2})\biggr]
\end{eqnarray} 
where, recall from section 3.1, $p_+$ is the positive light cone-momentum of the onium. So the $Y$ dependence of the Sudakov factors changes slightly. The second factor in~\eqref{eq:mueller} contained the probability that nothing happened after the last gluon is emitted at a rapidity $y$, which can be anywhere between $0$ and $Y$. This Sudakov factor is then given by, for a constant $\rho$
\begin{equation}
S=exp\biggl[-\bar{\bar{\alpha}}\int_y^Y dy\int_\rho d^2r_2\frac{r_{10}^2}{r_{12}^2r_{20}^2}\biggr]=exp\biggl[-2\bar{\alpha}\textrm{ln}\biggl(\frac{r_{10}}{\rho}\biggr)(Y-y)\biggr]
\end{equation}
Now one could think that we just do the same as we did above when we switch to a $y$ dependent cut-off. That is true but there is also a problem, we must know what $p_+$ is, now it won't be just given by the energy of the onium. This problem will also appear in the $x_2$ integral in~\eqref{eq:mueller}, $\rho (y)$ will now change in every emission and for every dipole we will have a different $\rho$. Therefore one would need to keep track of all the emissions during the evolution, this is not a problem in a MC simulation, after all that is what we have been doing in this thesis, but for a theoretical approach it is a major problem (for a constant $\rho$, these problems does not appear since it doesn't matter how many emissions there have been or which dipole is about to emit, the cut-off is the same anyway). It seems to be difficult to write down an equation for a generating functional as in~\eqref{eq:mueller}. However, it would be interesting if this was done. If one could just find such an equation, deriving a modified BK equation would not be difficult. Of course the new equation should not be too difficult to analyze, if one finds an equation that is not even numerically solvable it would not be so interesting.

\section{Conclusions and Summary}

Let us summarize the thesis. In other studies energy conservation has frequently been found to have a quantitatively large effect, and also to correspond to a significant part of NLO corrections. Our main objective has been to study saturation effects in DIS. To do this, we have constructed a Monte Carlo program which is based on Mueller's dipole formulation and respects energy conservation. Using our model we have studied $\lambda$, which is, in the BFKL approximation, defined by the relationship $\sigma \sim e^{\lambda y}$, and seen that there is a clear difference compared to previous studies, such as \cite{e9}, where energy conservation is not included. Throughout the thesis we have been working in the large $N_c$ limit. 

In onium-onium collisions we have seen that the difference between the one pomeron and the unitarised cross sections are quite small. It is hard to say if the cross section saturates but its growth with energy is reduced and we obtain $\lambda$ in the interval 0.10-0.15. In onium-nucleus collisions the difference between the unitarised and the one pomeron cross sections are larger, and the growth of the cross section slows down quite rapidly. It is also clearer in onium-nucleus scattering that ln$\sigma$ does not depend linearly on $Y$, except for small $Y$. The saturation sets in quite early and the growth of the cross section becomes smaller and smaller. In the onium-nucleus case however, one should be quite careful before drawing any conclusions. This process was not studied in detail in this thesis but it is going to be studied more deeply in future investigations. As we mentioned in the text the dipole distribution for the nucleus should be modified by taking into account normalization. The distribution we used was a very simple one and can be improved.    

In the second part of the thesis we studied the effects of dipole fusion processes. Using symmetry arguments, and in a factorizing approximation, we proposed an expression for the fusion factor, presented in~\eqref{eq:fusion4}. We used this expression in our Monte Carlo program, calculating cross sections for onium-onium collisions.  Due to lack of time we only studied the effects on the unitarised amplitude and without energy conservation. The value for the fusion parameter, $\xi$, cannot be determined by theoretical arguments however, and we have here considered various values for this parameter. Varying $\xi$ we obtained different values for $\lambda$, and we saw that fusion processes are not as important as energy conservation, in slowing down the growth of $\sigma$, at least for $Y\leqslant 14$. Before any definite conclusions are made though, the program should be tested more and it should be run with better statistics. It would also be interesting to study larger rapidity intervals, especially when including fusion processes. Fusion processes become interesting when the gluon density is high, which occurs at high rapidites. Studying larger $Y$ intervals is very time consuming however, especially for onium evolution without energy conservation. The problem is that the number of dipoles becomes very large as $Y$ increases, which limits the $Y$-range avaliable for numerical studies. 

We should also repeat that one needs to use a large $\xi$-values, larger than 10 at least, to see any significant effects of fusion processes. Since we derived~\eqref{eq:fusion4} in the large $N_c$ limit, which means that only neighboring dipoles can undergo fusion, the effects of fusion events are not so large for small and moderate $Y$, ($Y \lesssim 14$). Energy conservation has a larger effect on the growth of the cross section and its effects are seen earlier than fusion processes. Therefore energy conservation should have higher priority than fusion processes in DIS investigations.

\section{Acknowledgments}

I would like to express my gratitude to my supervisor G\"osta Gustafson who spent a lot of time answering my many questions, and from whom I certainly learned a lot. I would also like to thank Leif L\"onnblad for his, much appreciated, help with the Monte Carlo program. This thesis could not have been done without his help. Finally I would like to thank Torbj\"orn Sj\"ostrand for helping me with Gnuplot and my fellow masters students for keeping me good company.  

\appendix

\section{Appendix}

In this appendix we derive the formula~\eqref{eq:sann}. To find the $y$ distribution we use the formula~\eqref{eq:splittingweigth}. Remember that $\bold{r}_i=(0,0)$ and $\bold{r}_j=(1,0)$. Therefore we can write the splitting factor as
\begin{equation}
\bar{\bar{\alpha}}\int d^2\bold{r}_n \frac{r_{ij}^2}{r_{in}^2r_{jn}^2}=\bar{\bar{\alpha}}\int dr_xdr_y \frac{1}{(r_x^2+r_y^2)((r_x-1)^2+r_y^2)}=\int dr_xdr_y f(r_x,r_y)=F(y)
\label{eq:a1}
\end{equation}
The integration is done over all of $\mathbb{R}^2$ except for two circles with radius $\rho(y)$ where the first one is centered at $\bold{0}$ and the second one at $(1,0)$. The rapidity dependence in $F$ comes from the fact that $\rho=\rho (y)$. Using the veto algorithm the $y$ distribution is given by  
\begin{equation}
y=\bar{F}^{-1}(\bar{F}(y_i)-\textrm{ln}R)
\label{eq:veto}
\end{equation}
Here $\bar{F}$ is the primitive function of $F$, $y_i$ is the initial value of $y$ and $R$ is a random number. Finding $\bar{F}^{-1}$ is easy as long as $\rho$ is sufficiently small. But the integral in~\eqref{eq:a1} is difficult to evaluate when $\rho$ is not small and since we have a running $\rho$ we cannot always be sure that $\rho$ is sufficiently small. Therefore one needs to use an auxiliary function $g$, satisfying $g \geqslant f$. Before we do this, we note that $f(r_x,r_y)$ is symmetric around $r_x=1/2$. Therefore we do not have to consider the whole $\mathbb{R}^2$ plane but it is enough to look at the plane where $r_x<1/2$. We choose 
\begin{equation}
g(r_x,r_y)=\frac{2\bar{\bar{\alpha}}}{(r_x^2+r_y^2)(r_x^2+r_y^2+0.25)}
\end{equation}
We easily see that $g(r_x,r_y) \geqslant f(r_x,r_y)$ for all $(r_x,r_y)$ with $r_x <1/2$. Then we have
\begin{equation}
\int dr_xdr_yg(r_x,r_y)=16\pi \bar{\bar{\alpha}}\textrm{ln}\frac{\sqrt{\rho^2+0.25}}{\rho}=8\pi\bar{\bar{\alpha}}\textrm{ln}(1+\frac{e^{2y}p_+^2}{4})
\end{equation} 

We use $g(r_x,r_y)$ to generate the $r_x$ and $r_y$ values, which can be easily verified. Next we introduce a second auxiliary function, $h(y)$ given by
\begin{equation}
h(y)=8\pi\bar{\bar{\alpha}}\textrm{ln}(e^{2y}+\frac{e^{2y}p_+^2}{4}) \geqslant 8\pi\bar{\bar{\alpha}}\textrm{ln}(1+\frac{e^{2y}p_+^2}{4})=g(y)
\end{equation}
where the inequality follows from $y\geqslant 0$. Finding the primitive, $H$, of $h$ is easy, and it is also easy to find the inverse $H^{-1}$. Using~\eqref{eq:veto}, with $H$, we immediately obtain~\eqref{eq:rap}. To obtain the correct distribution we must accept the generated $r_x, r_y$ and $y$ values with the probability
\begin{equation}
\frac{f(r_x,r_y)}{g(r_x,r_y)}\frac{g(y)}{h(y)}
\label{eq:app1}
\end{equation}
We are not done however since we considered only the plane where $r_x <1/2$. When we generate the $r_x$, we reflect them around the symmetry axis, $r_x=1/2$, with a $0.5$ probability. Therefore instead of $g(r_x,r_y)$ we should in~\eqref{eq:app1} use 
\begin{equation}
g(r_x,r_y)=\frac{\bar{\bar{\alpha}}}{(r_x^2+r_y^2)(r_x^2+r_y^2+0.25)}+\frac{\bar{\bar{\alpha}}}{((r_x-1)^2+r_y^2)((r_x-1)^2+r_y^2+0.25)}
\end{equation}
Using this $g$, the correct formula is given by
\begin{equation}
\frac{f(r_x,r_y)}{g(r_x,r_y)}\frac{g(y)}{h(y)}=\frac{\textrm{ln}(1+\frac{e^{2y}p_+^2}{4})}{\textrm{ln}(e^{2y}+\frac{e^{2y}p_+^2}{4})}\biggl[\frac{1}{\frac{(r_x-1)^2+r_y^2}{r_x^2+r_y^2+0.25}+\frac{r_x^2+r_y^2}{(r_x-1)^2+r_y^2+0.25}}\biggr]
\end{equation}
which gives~\eqref{eq:sann}.

\end{document}